\documentclass[11pt,a4paper]{article}
\pdfoutput=1
\usepackage{jheppub}

\usepackage[normalem]{ulem}
\usepackage{graphicx}
\usepackage{epstopdf}
\usepackage{multirow}
\usepackage{rotating}
\usepackage{xcolor}

\def\vec#1{\mathbf{#1}}
\usepackage{wasysym}
\usepackage{slashed}
\usepackage[normalem]{ulem}

\newcommand{\mc}[1]{\ensuremath{\mathcal{#1}}}
\newcommand{\matrixel}[3]{\left< #1 \vphantom{#2#3} \right| #2 \left| #3 \vphantom{#1#2} \right>}
\newcommand{\avg}[1]{\left< #1 \right>}
\newcommand{\eq}[1]{\begin{equation} #1 \end{equation}}
\newcommand{\al}[1]{\begin{equation} \begin{aligned} #1 \end{aligned} \end{equation}}
\newcommand{\abs}[1]{\left| #1 \right|}

\def\ltorder{\mathrel{\raise.3ex\hbox{$<$}\mkern-14mu
    \lower0.6ex\hbox{$\sim$}}}
\def\gsim{\mathrel{\rlap{\lower4pt\hbox{\hskip1pt$\sim$}}
    \raise1pt\hbox{$>$}}}

\begin{document}
\title{The decay constants ${\mathbf{f_D}}$ and ${\mathbf{f_{D_{s}}}}$ in the continuum limit of ${\mathbf{N_f=2+1}}$ domain wall lattice QCD}

\author{
P.A.~Boyle$^a$,
L.~Del~Debbio$^a$,
A.~J\"uttner$^b$,
A.~Khamseh$^a$,
F.~Sanfilippo$^b$,
J.T.~Tsang$^b$\\[2mm]
RBC and UKQCD Collaborations}

\affiliation{$^a$Higgs Centre for Theoretical Physics, School of Physics \& Astronomy, University of Edinburgh, EH9 3FD, United Kingdom}
\affiliation{$^b$School of Physics and Astronomy, University of Southampton,  Southampton, SO17 1BJ, United Kingdom}

\abstract{
  We present results for the decay constants of the $D$ and $D_s$ mesons computed in lattice QCD with $N_f=2+1$ dynamical flavours.
  The simulations are based on RBC/UKQCD's domain wall ensembles with both physical and unphysical light-quark masses and lattice spacings in the range 0.11--0.07$\,$fm.
  We employ the domain wall discretisation for all valence quarks.

  The results in the continuum limit are $f_D=208.7(2.8)_\mathrm{stat}\left(^{+2.1}_{-1.8}\right)_\mathrm{sys}\,\mathrm{MeV}$ and $f_{D_{s}}=246.4(1.3)_\mathrm{stat}\left(^{+1.3}_{-1.9}\right)_\mathrm{sys}\,\mathrm{MeV}$ and $f_{D_s}/f_D=1.1667(77)_\mathrm{stat}\left(^{+57}_{-43}\right)_\mathrm{sys}$.
  Using these results in a Standard Model analysis we compute the predictions $|V_{cd}|=0.2185(50)_\mathrm{exp}\left(^{+35}_{-37}\right)_\mathrm{lat}$ and $|V_{cs}|=1.011(16)_\mathrm{exp}\left(^{+4}_{-9}\right)_\mathrm{lat}$ for the CKM matrix elements.}


\maketitle

\section{Introduction}\label{sec:Intro}
The charmed $D$ and $D_s$ mesons decay weakly into a lepton and a neutrino. 
The experimental measurement of the corresponding decay rates together with their prediction from within the Standard Model (SM) provides for a direct determination of the CKM matrix elements $|V_{cd}|$ and $|V_{cs}|$.
Leptonic $D_{(s)}$ decays have therefore been studied extensively  by a number of experiments (CLEO-c~\cite{Artuso:2005ym,Artuso:2007zg,Eisenstein:2008aa,Alexander:2009ux,Naik:2009tk,Ecklund:2007aa,Onyisi:2009th}, BES~\cite{Ablikim:2013uvu}, Belle~\cite{Zupanc:2013byn} and BaBar~\cite{delAmoSanchez:2010jg}).
Together with the perturbative prediction of electroweak contributions in the SM this leads to the results~\cite{PDG}
\begin{equation}\label{eq:expresults}
  |V_{cd}|f_{D^+}=45.91(1.05){\rm MeV}\,,\qquad |V_{cs}|f_{D_s^+}=250.9(4){\rm MeV}\,.
\end{equation}
The reliable SM prediction for the decay constants
\begin{equation}\label{eq:fPSdefinition}
  f_{D_q}=\frac{\matrixel{0}{\bar c\gamma_0\gamma_5 q}{D_q(0)}}{m_{D_q}},\qquad{\rm where\;}q=d,s\,,
\end{equation}
hence allows for the determination of $|V_{cd}|$ and $|V_{cs}|$ respectively. 
Combined with calculations of other CKM matrix elements~\cite{PDG,Aoki:2013ldr}, a number of SM tests of, for instance, the unitarity of the CKM matrix can be devised (see e.g. refs. \cite{Charles:2015gya,Bona:UTfit}).

Surprisingly, relatively few state-of-the-art predictions for the decay constants with a reliable control of systematic uncertainties exist to date (see e.g. the discussion of the results of $N_f=2+1$~\cite{Davies:2010ip,Bazavov:2011aa,Na:2012iu,Yang:2014sea} and $N_f=2+1+1$~\cite{Carrasco:2014poa,Bazavov:2014wgs} in the 2016 FLAG report~\cite{Aoki:2016frl}).
The computation of decay constants in lattice QCD is by now a well established exercise and many results with sub-percent precision within isospin symmetric QCD do exist for pions and kaons.
This is less so however in the case of the $D$- and $D_s$-meson decay constants.
The major difficulty there lies in the fact that with the charm-quark mass slightly above 1$\,\mathrm{GeV}$, cut-off effects arising from the charm mass remain a serious concern in lattice simulations.
Ensembles with dynamical quarks and sufficiently small lattice spacing to allow  controlled continuum extrapolations have only become feasible in recent years. 

Existing calculations try to deal with this in various ways by using discretisations tailored to reduce cut-off effects.
For instance, highly improved staggered quarks~\cite{Follana:2006rc}, the Fermilab approach~\cite{ElKhadra:1996mp}, overlap fermions~\cite{Neuberger:1997fp}, Osterwalder-Seiler fermions~\cite{Osterwalder:1977pc,Carrasco:2014poa} or non-perturbatively improved Wilson fermions~\cite{Juttner:2003ns,Heitger:2013oaa} are used as the charm-quark discretisation.

Here we present the first calculation of the $D$- and $D_s$-meson decay constants using the domain wall discretisation for the charm as well as the light and strange quarks on RBC/UKQCD's gauge ensembles with large volumes and physical values of the light-quark masses. 
Domain wall fermions (DWF) on the lattice provide chiral symmetry to a good approximation and as a result of this automatic $O(a)$ improvement.
In particular the latter is important when discretising charm quarks since no further work is required for tuning improvement coefficients in the action and for operators.
Discretising both the light as well as the charm quark within the same framework will also allow to correctly reproduce GIM cancellation~\cite{Glashow:1970gm} in lattice computations of quantities such as $\epsilon_K$~\cite{Christ:2014qwa}, $\Delta M_K$~\cite{Bai:2014cva} and processes such as $K\to\pi l^+ l^-$~\cite{Christ:2015aha,Christ:2016mmq} and $K\to\pi\nu\bar\nu$~\cite{Christ:2016eae} which the RBC/UKQCD collaboration is pursuing.

Given the novel nature of domain wall fermions as heavy quark discretisation we have investigated their properties in detail in two preparatory publications~\cite{Cho:2015ffa,quenched_sh}.
We studied the continuum limit behaviour of heavy-strange decay constants over a wide range of lattice cut-offs (2-6$\,\mathrm{GeV}$) within quenched QCD. In this way we determined parameters of the domain wall discretisation which resulted in small discretisation effects for charmed meson decay constants.
Refs. \cite{Cho:2015ffa,quenched_sh} show that DWF with suitably chosen parameters show mild cut-off dependence for charmed meson masses and decay constants.
We expect this to hold also in the presence of sea-quarks on RBC/UKQCD's gauge ensembles, thus the study at hand.

This work summarises our computation within this setup of the the $D$- and $D_s$-meson decay constants $f_D$ and $f_{D_s}$, respectively, their ratio and the Cabibbo-Kobayashi-Maskawa (CKM) matrix elements $\abs{V_{cd}}$ and $\abs{V_{cs}}$. 
For convenience we anticipate the numerical results:
{\begin{equation}\label{eq:allresults} \begin{aligned}
  f_D                &= 208.7(2.8)_\mathrm{stat}\left(^{+2.1}_{-1.8}\right)_\mathrm{sys}\,\mathrm{MeV},\\
  f_{D_s}             &= 246.4(1.9)_\mathrm{stat}\left(^{+1.3}_{-1.9}\right)_\mathrm{sys}\,\mathrm{MeV},\\
  \frac{f_{D_s}}{f_D} &= 1.1667(77)_\mathrm{stat}\left(^{+57}_{-43}\right)_\mathrm{sys}\\
  &\qquad\text{and}\\
  \abs{V_{cd}} &= 0.2185(50)_\mathrm{exp}(^{+35}_{-37})_\mathrm{lat},\\
  \abs{V_{cs}} &= 1.011(16)_\mathrm{exp}(^{+11}_{-\hphantom{0}9})_\mathrm{lat}\,,
\end{aligned}\end{equation}}
where errors are statistical, systematic and experimental, respectively.
Note that the quoted results for the decay constants are for isospin symmetric QCD.\footnote{See \cite{Carrasco:2015xwa} for a strategy to directly compute isospin breaking effects in meson decays.}
Systematic errors are due to choices made when fitting and parameterising the lattice data, lattice scale setting, finite volume, isospin breaking and renormalisation, based on $N_f=2+1$ flavour simulations.
Isospin breaking effects in the determination of CKM matrix elements are based on estimates.
In the determination of the CKM matrix element the lattice statistical and systematic errors have been combined in quadrature.

The paper is structured as follows: in section~\ref{sec:Setup} we summarise our numerical set up and give details of all simulation parameters.
Section~\ref{sec:Data Analysis} presents our complete analysis for the $D_{(s)}$-meson decay constants and constitutes the main body of this paper.
In section \ref{sec:CKMelements} we extract the corresponding CKM matrix elements before concluding in \ref{sec:summary}.

\section{Numerical simulations}\label{sec:Setup}
This report centres mainly around ensembles with physical light-quark masses in large volumes~\cite{RBCUKQCDPhysicalPoint}.  
The data analysis will however also take advantage of information obtained on ensembles with unphysically heavy pions~\cite{Allton:2008pn,Aoki:2010pe,Aoki:2010dy}.
The gauge field ensembles we use (cf. table \ref{tab:ensembles}) represent isospin symmetric QCD with $N_f=2+1$ dynamical flavours at three different lattice spacings in the range 0.11$\,\mathrm{fm}$-0.07$\,\mathrm{fm}$ ({\bf C}oarse, {\bf M}edium and {\bf F}ine). 
All ensembles have been generated with the Iwasaki gauge action~\cite{Iwasaki:1984cj,Iwasaki:1985we}.
For the discretisation of the quark fields we adopt either the DWF action with the M\"obius kernel~\cite{Brower:2004xi,Brower:2005qw,Brower:2012vk} or the Shamir kernel~\cite{Kaplan:1992bt,Shamir:1993zy}.
The difference between both kernels in our implementation corresponds to a rescaling such that M\"obius domain wall fermions (MDWF) are loosely equivalent to Shamir domain wall fermions (SDWF) at twice the extension in the fifth dimension~\cite{RBCUKQCDPhysicalPoint}. 
M\"obius domain wall fermions are hence cheaper to simulate while providing the same level of lattice chiral symmetry.
Results from both formulations of domain wall fermions lie on the same scaling trajectory towards the continuum limit with cut-off effects starting at $O(a^2)$. 
Even these $O(a^2)$ cut-off effects themselves are expected to agree between our M\"obius and Shamir formulations, with their relative difference at or below the level of 1\%~\cite{RBCUKQCDPhysicalPoint} for the finite values of $L_s$ used in our simulations.
\begin{table}
  \begin{center}
    \begin{tabular}{c c c c c c c c c}
      \hline\hline\\[-4mm]
      Name &  DWF & $L/a$ & $T/a$ &  $a^{-1}[\mathrm{GeV}]$ & $m_\pi[\mathrm{MeV}$] & hits/conf & confs & total\\\hline
      C0   & MDWF & 48    & 96     & 1.7295(38)   & 139.15(36)   & 48 & 88  & 4224 \\
      C1   & SDWF & 24    & 64     & 1.7848(50)   & 339.789(12)  & 32 & 100 & 3200 \\
      C2   & SDWF & 24    & 64     & 1.7848(50)   & 430.648(14)  & 32 & 101 & 3232 \\\hline
      M0   & MDWF & 64    & 128    & 2.3586(70)   & 139.35(46)   & 32 & 80  & 2560 \\
      M1   & SDWF & 32    & 64     & 2.3833(86)   & 303.248(14)  & 32 & 83  & 2656 \\
      M2   & SDWF & 32    & 64     & 2.3833(86)   & 360.281(16)  & 16 & 77  & 1232 \\\hline
      F1   & MDWF & 48    & 96     & 2.774(10)\hphantom{0} & 234.297(10)  & 48 & 82  & 3936 \\
      \hline \hline
    \end{tabular}
  \end{center}
  \caption{
    This table summarises the main parameters of our $N_f=2+1$ ensembles.
    C stands for coarse, M for medium and F for fine, MDWF for M\"obius and SDWF for Shamir DWF.}
  \label{tab:ensembles}
  \end{table}

\begin{table}
  \begin{center}
    \begin{tabular}{c@{\hspace{2mm}} c@{\hspace{2mm}} c@{\hspace{2mm}} c@{\hspace{2mm}} c@{\hspace{2mm}} c@{\hspace{2mm}} c@{\hspace{2mm}} c@{\hspace{2mm}} c}
      \hline\hline\\[-4mm]
      Name & DWF & $M_5$  & $L_s$ & $am_l^\mathrm{uni}$ & $am_s^{\mathrm{uni}}$ & $am_s^{\mathrm{sim}}$ & $am_s^{\mathrm{phys}}$ & $\Delta m_s/m_s^{\mathrm{phys}}$\\\hline
      C0  & MDWF    & 1.8    &  24   & 0.00078  & 0.0362   & 0.0362           & 0.03580(16) & 0.0112(45) \\
      C1  & SDWF    & 1.8    &  16   & 0.005    & 0.04     & 0.03224, 0.04    & 0.03224(18) & -\\
      C2  & SDWF    & 1.8    &  16   & 0.01     & 0.04     & 0.03224          & 0.03224(18) & - \\\hline
      M0  & MDWF    & 1.8    &  12   & 0.000678 & 0.02661  & 0.02661          & 0.02539(17) & 0.0476(70) \\
      M1  & SDWF    & 1.8    &  16   & 0.004    & 0.03     & 0.02477, 0.03    & 0.02477(18) & - \\
      M2  & SDWF    & 1.8    &  16   & 0.006    & 0.03     & 0.02477          & 0.02477(18) & - \\\hline
      F1  & MDWF    & 1.8    &  12   & 0.002144 & 0.02144  & 0.02144          & 0.02132(17) & -0.0056(80) \\
      \hline \hline
    \end{tabular}
  \end{center}
  \caption{
    Domain wall parameters for the light and strange quarks.
    All quoted values for $am_l$ and $am_s$ are bare quark masses in lattice units.
    The column DWF corresponds to the chosen domain wall fermion formulation where `MDWF' corresponds to M\"obius domain wall fermions, `SDWF' to Shamir domain wall fermions.
    All light quarks are simulated at their unitary value $am_l^\mathrm{uni}$.
    Valence strange quarks were simulated at $am_s^\mathrm{sim}$.
    Note that the value of the physical strange-quark mass $am_s^\mathrm{phys}$ slightly disagrees with the unitary strange-quark mass $am_s^\mathrm{uni}$.}
  \label{tab:DWFls}
\end{table}

Basic properties of all ensembles used in this work are summarised in tables \ref{tab:ensembles} and \ref{tab:DWFls}.
The lattice scale and physical light-quark masses have been determined for all ensembles bar F1 in~\cite{RBCUKQCDPhysicalPoint} using $m_\pi$, the $m_K$ and the $m_\Omega$ as experimental input. 
The finest ensemble F1 was generated later as part of RBC--UKQCD's charm and bottom physics program and its properties are described in appendix~\ref{app:ensemble F1}.
We repeated the analysis of~\cite{RBCUKQCDPhysicalPoint} after including also ensemble F1 and in this way determined the value of the lattice spacing and physical strange-quark mass on this ensemble.

In the valence sector we simulate light, strange and charm-like heavy quarks.
The light-quark masses were chosen to be unitary.
The strange-quark mass was slightly adjusted (partially quenched) to its physical value as determined in~\cite{RBCUKQCDPhysicalPoint}, in cases where this was known prior to running the measurements (C1, C2, M1 and M2).
Otherwise the unitary value was chosen.
The parameters of the light and strange sector are listed in table \ref{tab:DWFls}.

Besides the bare quark mass, DWF have two further input parameters that need to be specified in each simulation: the extent of the fifth dimension $L_s$ and the {\it domain wall height} parameter $M_5$, respectively (for details see \cite{RBCUKQCDPhysicalPoint,quenched_sh}).
More specifically, $M_5$ is the negative mass parameter in the 4-dimensional Wilson Dirac operator which resides in the 5-dimensional DWF Dirac operator.
The parameter $M_5$ can have some effect on both, the rate of exponential decay of the physical modes away from the boundary in the 5th dimension and the energy scale of unphysical modes that are not localised at the boundary.
While changing $L_s$ mostly changes the magnitude of residual chiral symmetry breaking, the choice of $M_5$ in principle changes the ultraviolet properties of the discretisation.
Hence, different choices of $M_5$ lead to in principle different scaling trajectories.
The most important consequence is that calculations combining (quenched) charm quarks with a distinct $M_5$ from the light sea quarks must formally be treated as a mixed action when renormalising flavour off diagonal operators.

Two observations in this context which we made in our quenched DWF studies~\cite{Cho:2015ffa,quenched_sh} are crucial for understanding the choice of simulation parameters made here:
studying the pseudoscalar heavy-heavy and strange-heavy decay constants we found cut-off effects to be minimal for $M_5\approx1.6$. 
We find that the residual chiral symmetry breaking effects are well suppressed for $L_s=12$.
At the same time we observed a rapid increase in discretisation effects as the bare input quark mass was increased above $am_h\approx 0.4$.
Here we  work under the assumption that these observations carry over to the case of dynamical simulations with $N_f=2+1$ flavours which motivates our choice of $M_5=1.6$ for charm DWF keeping $am_h\le 0.4$.
As we will see later this assumption is well justified.
For the light quarks we use $M_5=1.8$ both in the valence- and sea-sector.

All simulated bare charm-quark masses are listed in table \ref{tab:DWFh}.
Note, that we allow for one exception to the bound $am_h\le 0.4$ by generating data for $am_h=0.45$ on ensemble C0.
With this we tested whether the reach in the heavy quark mass for DWF with $M_5=1.6$ observed in the quenched theory~\cite{quenched_sh} also persists in the dynamical case.
The quantity that we monitor in this context is the {\it residual quark mass} $am_{\rm res}$, which provides an estimate of residual chiral symmetry breaking in the DWF formalism.
It is defined in terms of the axial Ward identity (AWI)
\begin{equation}
  \label{mobpcac}
  a \Delta^-_\mu \langle P(x){\cal A}_\mu(y) \rangle = \langle P(x)\left[2am P(y) + 2 J_{5q}(y)\right] \rangle \,,
\end{equation}
where $\Delta_\mu^-$ is the lattice backward derivative, $am$ the bare quark mass in lattice units in the Lagrangian, $P$ is the pseudoscalar density and $J_{5q}$ is the pseudoscalar density in the centre of the 5th dimension.
It motivates the definition
\begin{equation}
  \label{eq:residual mass}
  am_{\rm res}=\frac{\sum\limits_{\mathbf{ x}} \langle J_{5q}(x) P(0)\rangle}{\sum\limits_{\mathbf{ x}}\langle P(x) P(0)\rangle}\,.
\end{equation}

\begin{figure}
  \center
  \includegraphics[width=0.8\textwidth]{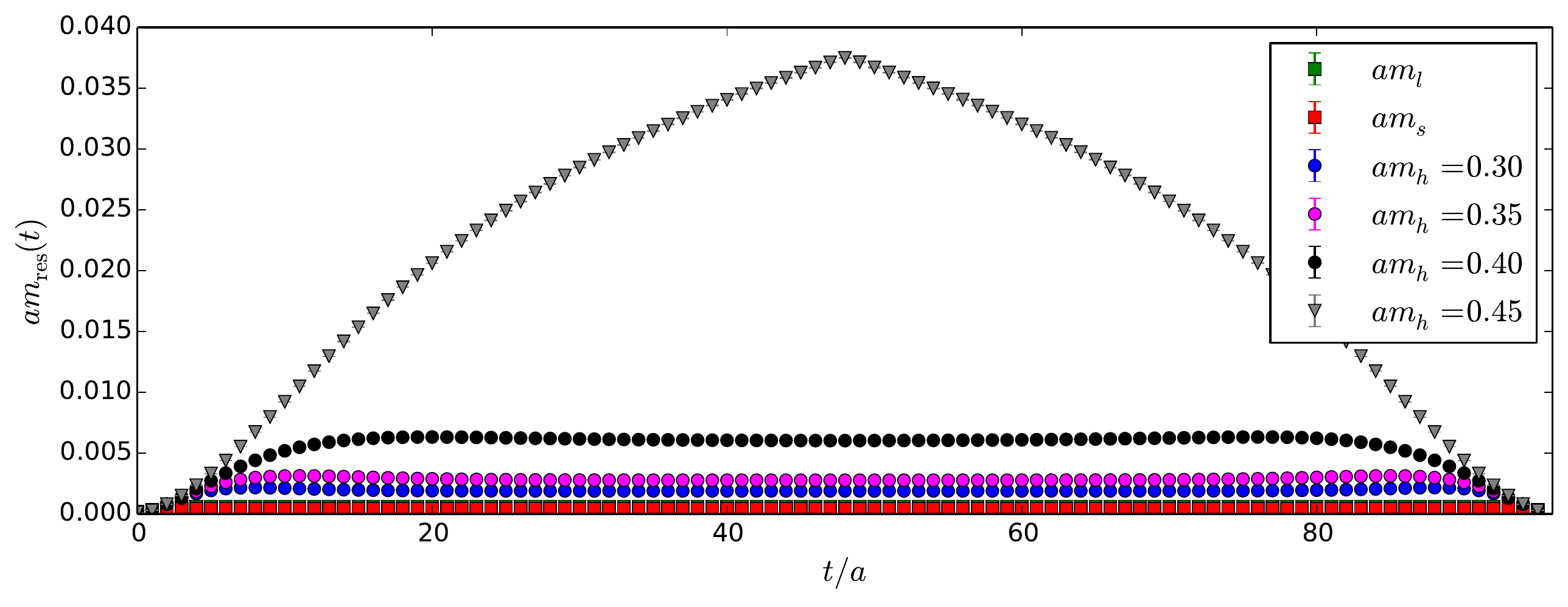}
  \caption{The effective residual mass on the C0 ensemble for all simulated masses.} 
  \label{fig:mres_C0}
\end{figure}

Figure \ref{fig:mres_C0} shows the behaviour of the residual mass on the C0 ensemble.
As expected (see reference \cite{quenched_sh} for details) the residual mass does plateau and remains flat for $am_h \lesssim 0.4$, confirming the validity and upper bound of the mass point at $am_h=0.4$.
The fact that this behaviour is not observed with $am_h=0.45$ indicates that cut-off effects change in nature when $am_h$ is pushed beyond 0.4, in agreement with our findings in ref. \cite{quenched_sh}.
We therefore exclude any data with $am_h>0.4$ in the remainder of this paper.

\begin{table}
  \begin{center}
    \begin{tabular}{c c r c c}
      \hline\hline\\[-4mm]
      Name & $M_5$&$L_s$ & $am_h^{\mathrm{bare}}$         \\\hline
      C0   & 1.6    &  12   & 0.3,  0.35, 0.4, {\color{red} 0.45}  \\
      C1   & 1.6    &  12   & 0.3,  0.35, 0.4             \\
      C2   & 1.6    &  12   & 0.3,  0.35, 0.4             \\\hline
      M0   & 1.6    &   8   & 0.22, 0.28, 0.34, 0.4       \\
      M1   & 1.6    &  12   & 0.22, 0.28, 0.34, 0.4       \\
      M2   & 1.6    &  12   & 0.22, 0.28, 0.34, 0.4       \\\hline
      F1   & 1.6    &  12   & 0.18, 0.23, 0.28, 0.33, 0.4 \\
      \hline \hline
    \end{tabular}
  \end{center}
  \caption{
    M\"obius domain wall parameters for the heavy quarks on all ensembles.
    All quoted values for $am_h$ are bare quark masses in lattice units.
    As described in the text, the value indicated in red was only used to verify our assumptions about the applicability of the quenched pilot study to the dynamical case.}
  \label{tab:DWFh}
\end{table}

\section{Data Analysis}\label{sec:Data Analysis}
In this section we describe how we make predictions for the decay constants starting from the evaluation of Euclidean two-point correlation functions on all ensembles and for all parameters discussed above.
In particular, from fits to the simulated data we determine pseudoscalar masses $m_P$ and decay constants $f_P$ and ratios thereof, which depend on the simulation parameters $(a,m_l,m_h,m_s)$. Of particular relevance for the analysis are $P=\pi,\, K,\, D,\,D_s$ and $\eta_c$ (the latter one corresponding to an unphysical $c\bar{c}^\prime$ state made of two distinct flavours of quarks having the same charm quark mass).
We extrapolate the data for each observable $\mc{O}$ to physical light, strange and charm-quark masses as well as to vanishing lattice spacing and infinite volume, $\mathcal{O}(a=0,m_l^\mathrm{phys},m_h^\mathrm{phys},m_s^\mathrm{phys})$.

Besides the decay constants and ratios thereof, for reasons that will become clear later, we will carry out the analysis of the decay constants in terms of the quantity $f_P\sqrt{m_P}$ and only remove the factor $\sqrt{m_P}$ in the final step.

\subsection{Correlation functions} \label{subsec:corrfits}

In practice we determine the matrix element in equation (\ref{eq:fPSdefinition}) from the time dependence of Euclidean QCD zero-momentum two-point correlation functions,
\begin{equation}
  C_{i,MN}(t) \equiv \sum_{\vec{x,y}} \langle
  \,O_{i,M}(t,\vec{y})\, \left(O_{i,N}\right)^\dagger(0,\vec{x})\,\rangle
  = \sum\limits_k
  \frac{Z^{(k)}_{i,M}\left( Z^{(k)}_{i,N}\right)^\ast}{2E^{(k)}_i} \left(e^{-E^{(k)}_it}\pm e^{-E^{(k)}_i(T-t)}\right)\, .
  \label{eq:twopt}
\end{equation}
We consider the cases $i=\pi,K,D,D_s$ or $\eta_{c}$.
The operators $O_{i,M}$ are interpolating operators with the quantum numbers of the desired mesons, e.g. $O_{D_s,M}= \bar c\, \Gamma_M s$\,, where we consider $\Gamma_A=\gamma_0\gamma_5$ and $\Gamma_P=\gamma_5$, respectively.
The sum on the r.h.s. of eq.~\eqref{eq:twopt} is over excited states $k$ and in practice we will consider the ground and first excited state in the data analysis.
The constants $Z_{i,M}^{(k)}$ are defined by $Z_{i,M}^{(k)}=\langle\,P_i^{(k)}\,|\,\left( O_{i,M}\right)^\dagger\,|\,0\,\rangle$ where $P_{i}^{(k)}$ is the corresponding $k$th excited meson state.

When computing the quark propagators we use $\mathbb{Z}(2) \times \mathbb{Z}(2)$ stochastic wall sources~\cite{Foster:1998vw,McNeile:2006bz,Boyle:2008rh} on a large number of time planes.
Details of how many different source planes are used for the various ensembles are listed in the column ``hits/conf'' in table \ref{tab:ensembles}.
The results on a given gauge configuration are averaged into one bin.

The calculation of the light and strange quark propagators were performed using the HDCG algorithm~\cite{Boyle:2014rwa}, reducing the numerical cost and hence making this computation feasible.
For the heavy quark propagators a CG inverter was used and we monitored satisfactory convergence using the time-slice residual introduced in ref. \cite{Juttner:2005ks}.

Masses and decay constants have been determined by simultaneous multi-channel fits to the two-point correlation functions $C_{i,AP}$ and $C_{i,PP}$ for a given choice of $i$. We attempted the use of correlated fits, but found the correlation matrix to be too poorly estimated for a reliable inversion. We therefore carry out uncorrelated fits, i.e. assume the correlation matrix to be diagonal (compare ref. \cite{RBCUKQCDPhysicalPoint}).

The statistical precision of the ground state mass and matrix elements is improved by fitting the ground state as well as the first excited state, allowing for earlier time slices (with smaller statistical errors) to contribute to the fit.
This was done for $i=D,D_s$, where we are interested in the matrix elements and not merely in the masses as in the case for $i=\pi,\eta_c$.
During all these fits the $\chi^2/\mathrm{d.o.f.}$ were monitored.
Tables with the bare results in lattice units on all ensembles can be found in appendix~\ref{app:correlatorfits}.

\begin{figure}
  \center
  \includegraphics[width=0.49\textwidth]{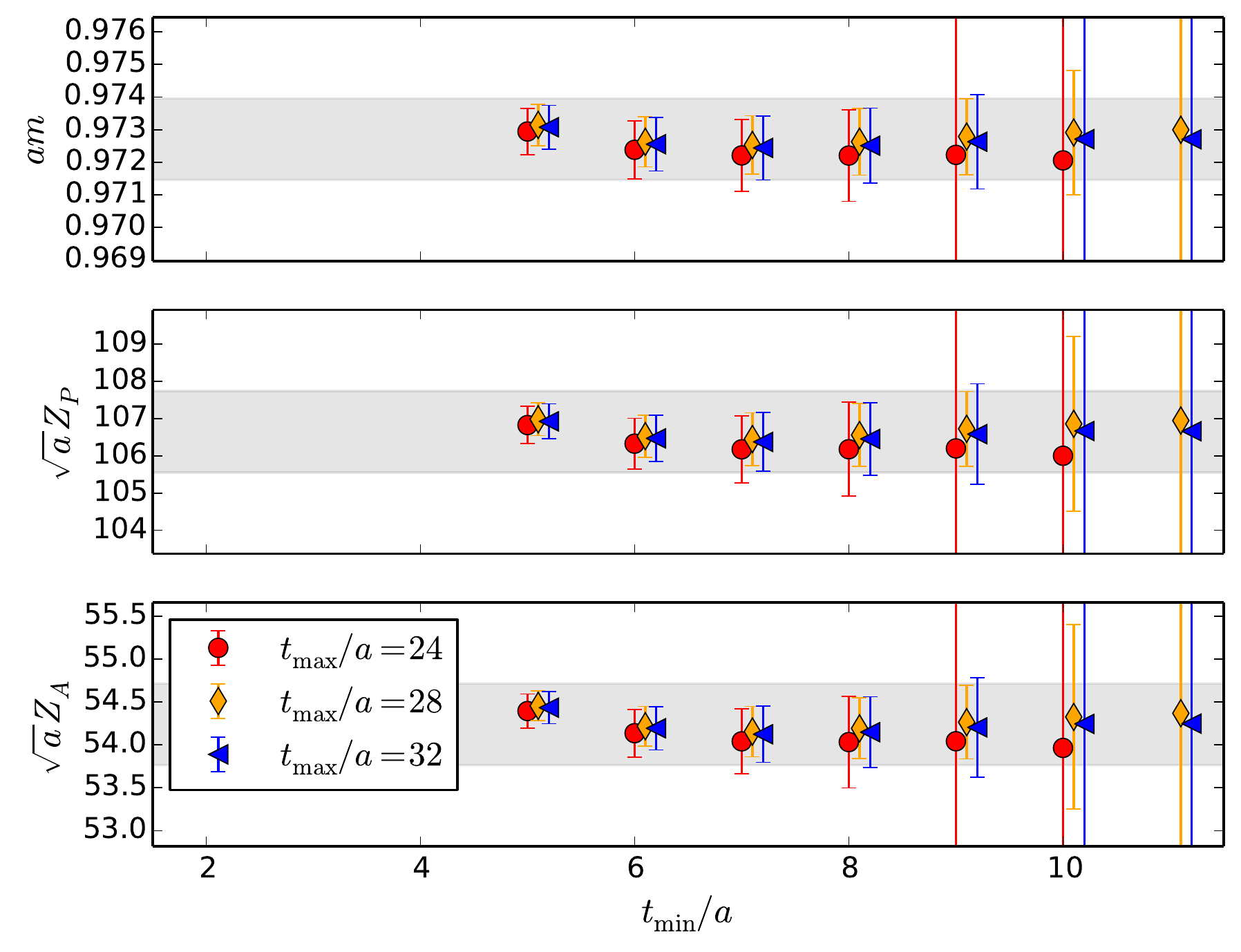}
  \includegraphics[width=0.49\textwidth]{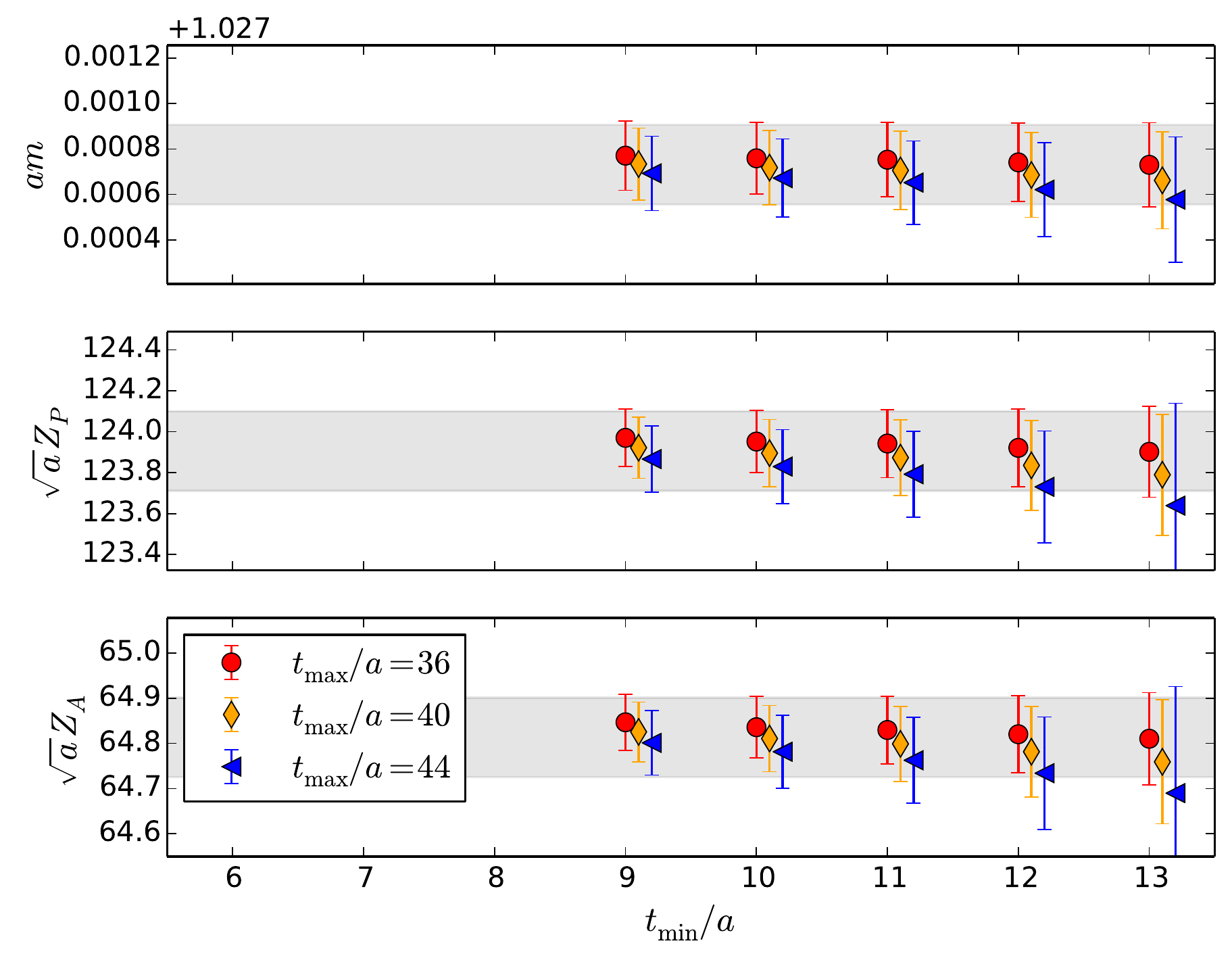}
  \caption{
    Variation of the time slices included in the correlation function fits for a heavy-light (left) and heavy-strange (right) meson on the C0 ensemble.
    The grey errorband shows the result of the conservatively chosen fit with $t/a\in[8,30)$ (left) and $t/a\in[12,37)$ (right).
        The large errorbars for large values of $t_\mathrm{min}$ arise when the excited state can no longer be resolved.} 
  \label{fig:fitrangevariation}
\end{figure}

Figure \ref{fig:plateaux} illustrates the correlator fit for the heaviest charm-quark mass on the C0 ensemble.
\begin{figure}
  \begin{center}
    \includegraphics[width=0.49\textwidth]{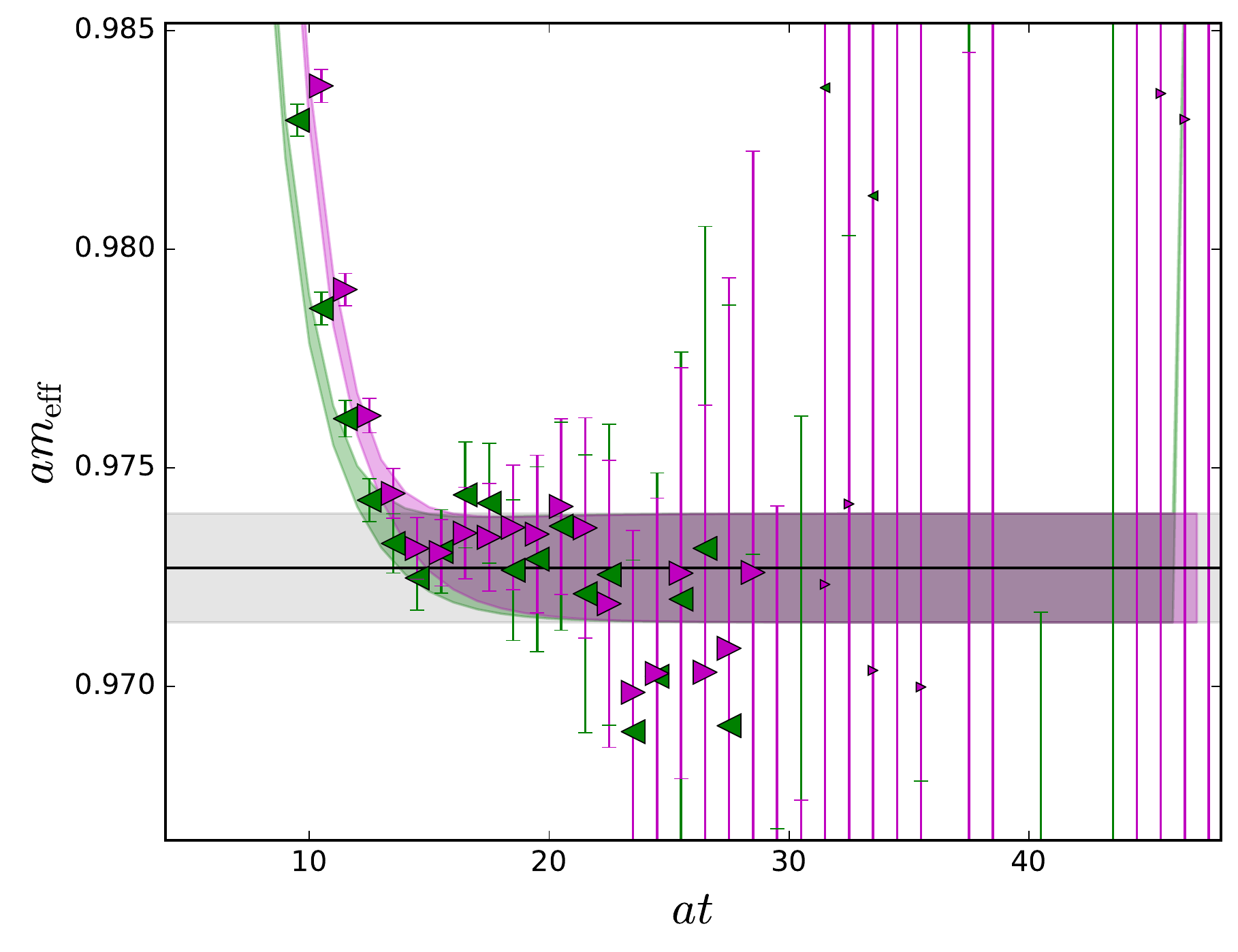}
    \includegraphics[width=0.49\textwidth]{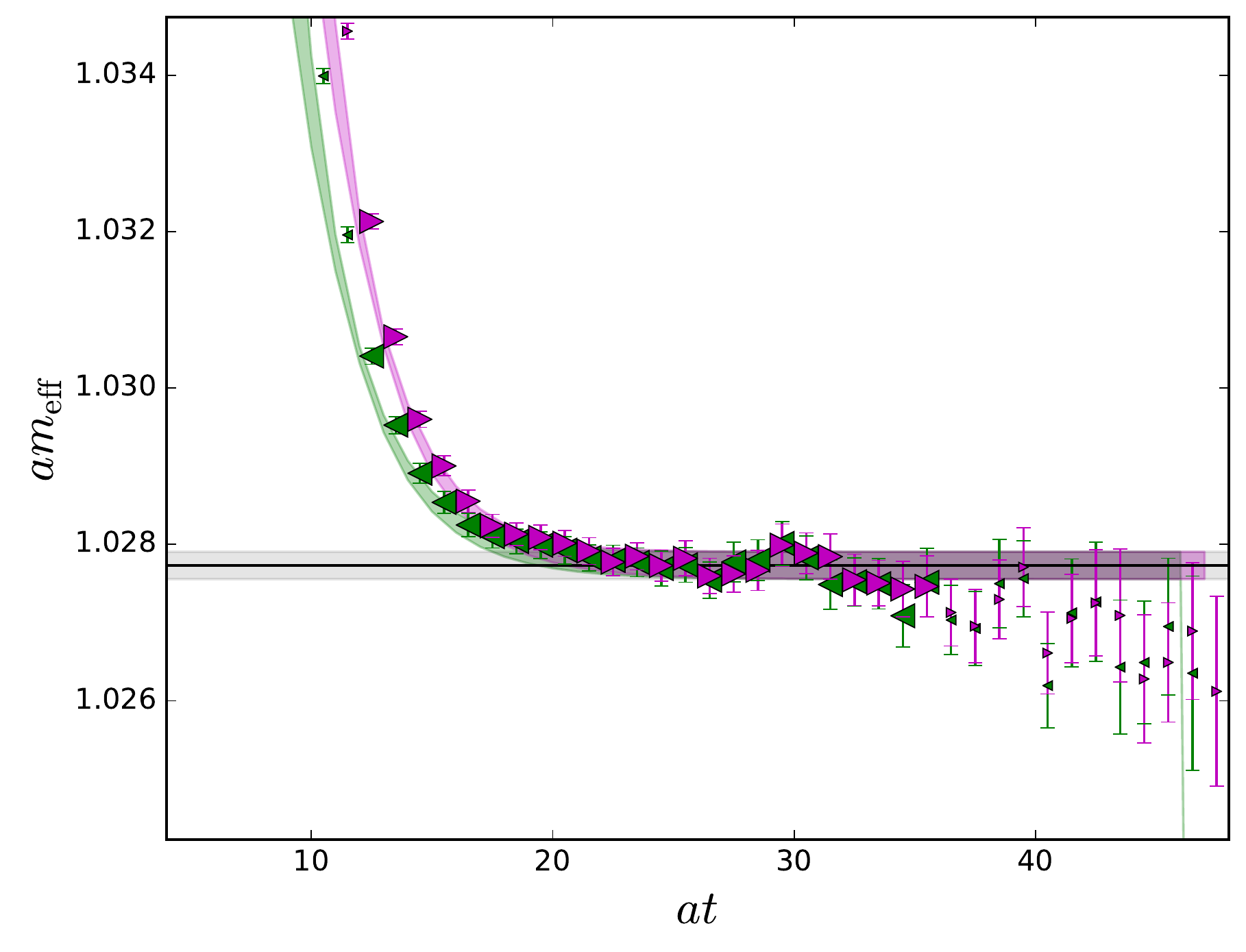}
  \end{center}
  \caption{
    Example of excited state correlation function fits for a heavy-light (left) and heavy-strange (right) meson on the C0 ensemble.
    The black line and grey shaded region is the fit result for the ground state mass.
    The green left-facing and magenta right-facing triangles show the effective mass as obtained from the simulated data for $\avg{AP}$ and $\avg{PP}$, respectively.
    Larger symbols correspond to data points that enter the fit.
    The coloured shaded regions are the fit result for the $\avg{AP}$ and $\avg{PP}$ correlation functions respectively.} 
  \label{fig:plateaux}
\end{figure}
The fit ranges were chosen by systematically varying $t_\mathrm{min}$ and $t_\mathrm{max}$, fixing them in a region where no dependence is observed.
Figure \ref{fig:fitrangevariation} shows the ground state results of varying $t_\mathrm{min}$ and $t_\mathrm{max}$ for the case of the heaviest mass point on the coarse ensemble C0.

\begin{figure}
  \center
  \includegraphics[width=\textwidth]{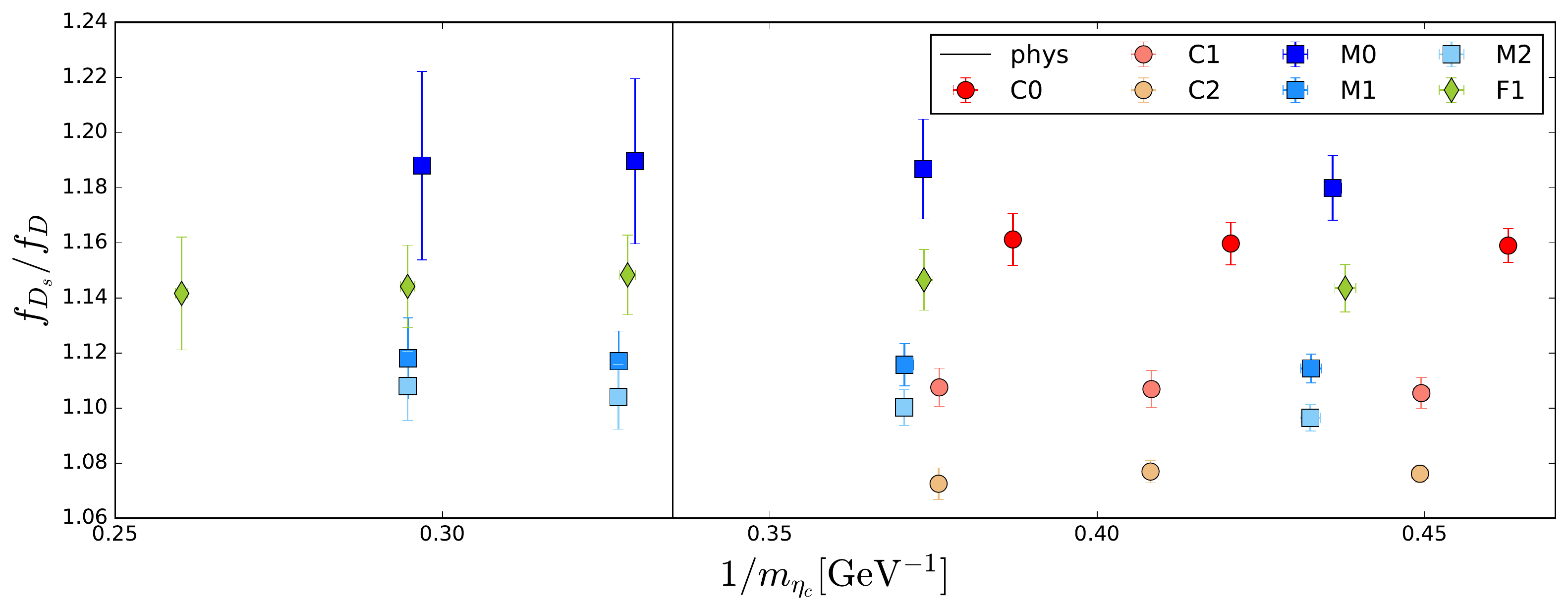}
  \caption{
    Collected data for the ratio of decay constants $f_{D_s}/f_D$ as obtained from the correlator fits.
    The solid vertical line labelled ``phys'' corresponds to the physical $\eta_c$ mass.}
  \label{fig:datacollection}
\end{figure}

A first impression of the range of ensembles (lattice spacing, light sea quark mass, charm-quark mass) for which we generated data is given in figure~\ref{fig:datacollection}, representatively for the ratio of decay constant $f_{D_s}/f_D$ plotted against the inverse of the measured $\eta_{c}$ mass.

\subsection{Non perturbative renormalisation} \label{subsec:npr}
To make contact between lattice regulated data and quantities in a continuum theory, the heavy-light current needs to be renormalised.
Since we use a mixed action current, by using a different value of $M_5$ for the light and strange quarks ($M_5=1.8$) to the heavy quarks ($M_5=1.6$), the usual domain wall axial Ward identity eq.~\eqref{mobpcac} is not satisfied.

Since in the free field theory the modification of the action represents a modest change to irrelevant parameters, we might hope the impact on renormalisation constants is small.
This is something we can verify.
In fact, the estimate of the systematic error arising from this change in action is found to be below percent level, and this is discussed in more detail in section \ref{subsubsec:NPRstudy} along with a proposal for the determination of fully non-perturbatively renormalised axial currents using appropriate ratios of off-shell mixed and unmixed action vertex functions.

Since empirically this is indeed a small effect, in all of the following we will extract the renormalisation constants from a light-light unmixed action current and associate a systematic error devised from the non-perturbative renormalisation (NPR) study. 

It is worth mentioning that we have recently developed a massive renormalisation scheme~\cite{Boyle:2016wis}, RI/mSMOM, by extending the massless RI/SMOM scheme~\cite{Martinelli:1994ty,Sturm:2009kb}, with the  renormalised composite fields defined away from the chiral limit.
This includes finite masses in one, i.e. the mixed case, or both quark fields entering the bilinear operator.
Using this scheme the renormalisation constant for the heavy-light axial current can be extracted non-perturbatively.
For more details, refer to ref. \cite{Boyle:2016wis} and appendix~\ref{app:nprsmom}.

\subsubsection{Unmixed action axial current renormalisation constants}
The light-light axial renormalisation constant can be found from fitting the time behaviour of the relation
\eq{
  Z_A^\mathrm{eff}(t) = \frac{1}{2} \left[ \frac{C_{\mc{A}P}(t-1/2)+C_{\mc{A}P}(t+1/2)}{2C_{AP}(t)} + \frac{2 C_{\mc{A}P}(t+1/2)}{C_{AP}(t-1)+  C_{AP}(t+1)} \right] 
  \label{eq:ZAeff}
}
to a constant~\cite{Boyle:2015vda,RBCUKQCDPhysicalPoint}. Here $C_{\mc{A}P}(t)$ is the correlation function between the conserved point-split axial vector current defined on the links between the lattice sites~\cite{Boyle:2015vda} and the pseudoscalar density, whilst $C_{AP}(t)$ is the same correlation functions with the conserved axial vector current replaced by the local current one defined on the lattice sites.

\begin{figure}
  \begin{center}
    \includegraphics[width=\textwidth]{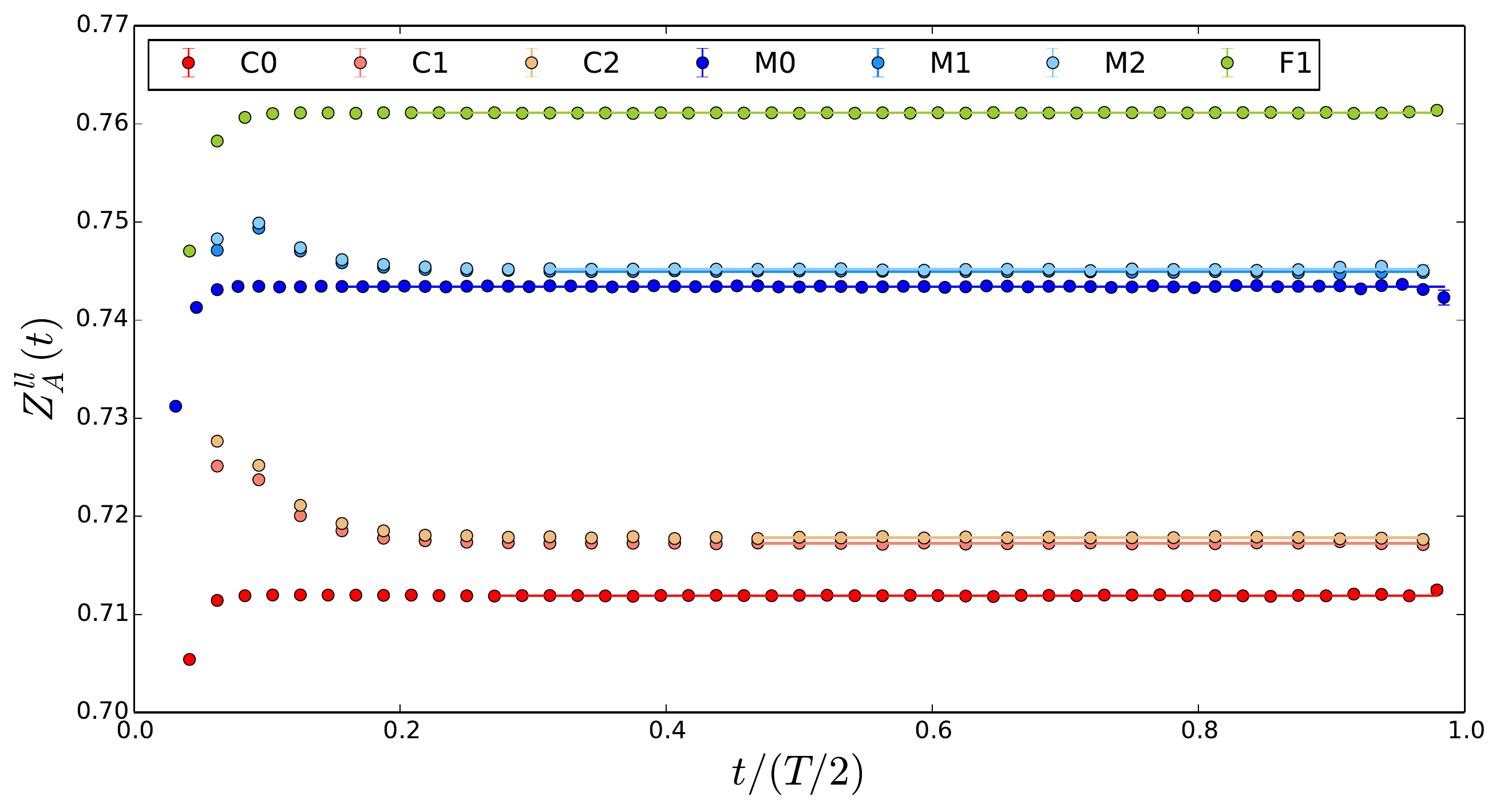}
  \end{center}
  \caption{
    The time behaviour of $Z_A^\mathrm{eff}(t)$ scaled to the interval $[0,1]$ for all ensembles.
    The solid lines correspond to the fit results obtained by fitting a constant to the plateau region of the data.
    The fit results are summarised in table \ref{tab:renormalisation_constants}.}
  \label{fig:ZA_all}
\end{figure}

The folded time behaviour of the light-light current for all ensembles scaled to the interval $[0,1)$ is shown in figure \ref{fig:ZA_all}.
As expected we can see a lattice spacing dependence.
The slight difference between C1 and C2 (M1 and M2) arises from the slightly different light-quark mass $m_l$.
We can also identify a plateau region to which we can fit a constant to obtain the values of the renormalisation constants.
The results of these fits are listed in table \ref{tab:renormalisation_constants} and are in good agreement with ref. \cite{RBCUKQCDPhysicalPoint}.
  
\begin{table}
  \center
  \begin{tabular}{|c|c|}
    \hline
    ens & $Z_A^{ll}$ \\
    \hline
    C0 &  0.711920(24) \\
    C1 &  0.717247(67) \\
    C2 &  0.717831(53) \\\hline
    M0 &  0.743436(16) \\
    M1 &  0.744949(39) \\
    M2 &  0.745190(40) \\\hline
    F1 &  0.761125(19) \\\hline
  \end{tabular}
  \caption{The values for $Z_A$ for the various ensembles found from fitting a constant to the time dependence shown in figure \ref{fig:ZA_all}.}
  \label{tab:renormalisation_constants}
\end{table}

\subsubsection{Vertex functions of mixed action current}\label{subsubsec:NPRstudy}
We use the non-exceptional Rome-Southampton renormalisation scheme, RI/SMOM,~\cite{Martinelli:1994ty,Sturm:2009kb} to investigate the effects of a change in the action on quantities entering the axial current renormalisation constant.
In particular we evaluate the variation of the projected, amputated vertex function for the axial current $\mathcal{P}[\Lambda_\text{A}]$, on each of the ensembles C2, M1 and F1.

The details of the numerical computation of the amputated axial vertex function $\Lambda_\text{A}$ in the RI/SMOM scheme are discussed in appendix~\ref{app:nprsmom}.
Tables \ref{tab:lambdaAratiosC2}, \ref{tab:lambdaAratiosM1} and \ref{tab:lambdaAratiosF1} present the ratios of $\mathcal{P}[\Lambda_\text{A}]$, for different combinations of actions i.e. $(M_5^1,M_5^2)=(1.8,1.8), (1.6,1.8), (1.6,1.6)$ at around 2$\,$GeV.
In all cases the unitary light-quark mass - which is assumed to be sufficiently close to the chiral limit - is used.
The data has been generated using ten gauge field configurations which leads to sufficiently precise results.
We see that the ratio $\frac{\mathcal{P}[\Lambda_\text{A}](1.8,1.8)}{\mathcal{P}[\Lambda_\text{A}](1.6,1.6)}$ on each of the ensembles has the largest deviation from unity as compared to the other ratio combinations, which is expected since both the quark fields entering the bilinear have different actions between the numerator and the denominator.

The main feature emerging from this study is that the deviation from unity is at most of order $0.4\%$ across a range of momenta around 2 GeV.
This is negligible on the scale of our other uncertainties and for the purposes of the present work we can simply include it as a sub-dominant systematic error.

\begin{table}
  \begin{center}
    \begin{tabular}{cccccc}
      \hline\hline\\[-4mm]
      $(ap)^2$ & $p\,[\mathrm{GeV}]$ & $\frac{\mathcal{P}[\Lambda_\text{A}](1.8,1.8)}{\mathcal{P}[\Lambda_\text{A}](1.6,1.6)}$ & $\frac{\mathcal{P}[\Lambda_\text{A}](1.6,1.8)}{\mathcal{P}[\Lambda_\text{A}](1.6,1.6)}$ & $\frac{\mathcal{P}[\Lambda_\text{A}](1.8,1.8)}{\mathcal{P}[\Lambda_\text{A}](1.6,1.8)}$\\\hline
      1.037 & 1.817 & 0.996816(35) & 0.998149(32) & 0.998664(12)\\
      1.133 & 1.900 & 0.996878(41) & 0.998180(33) & 0.998695(14)\\
      1.234 & 1.982 & 0.996943(37) & 0.998220(29) & 0.998721(17)\\
      1.339 & 2.065 & 0.997009(31) & 0.998263(23) & 0.998743(17)\\
      1.448 & 2.148 & 0.997084(28) & 0.998312(20) & 0.998770(15)\\
      \hline \hline 
    \end{tabular}
  \end{center}
  \caption{
    The ratios of projected amputated vertex function for the axial currents with different actions on the C2 ensemble.
    The quark mass for both fields is taken to be $am_l=0.01$. }
  \label{tab:lambdaAratiosC2}
\end{table}

\begin{table}
  \begin{center}
    \begin{tabular}{cccccc}
      \hline\hline\\[-4mm]
      $(ap)^2$ & $p\,[\mathrm{GeV}]$ & $\frac{\mathcal{P}[\Lambda_\text{A}](1.8,1.8)}{\mathcal{P}[\Lambda_\text{A}](1.6,1.6)}$ & $\frac{\mathcal{P}[\Lambda_\text{A}](1.6,1.8)}{\mathcal{P}[\Lambda_\text{A}](1.6,1.6)}$ & $\frac{\mathcal{P}[\Lambda_\text{A}](1.8,1.8)}{\mathcal{P}[\Lambda_\text{A}](1.6,1.8)}$\\\hline
     0.583 & 1.820 & 0.996774(55) & 0.998139(53) & 0.998508(24)\\
     0.637 & 1.903 & 0.996805(66) & 0.998132(45) & 0.998516(32)\\
     0.694 & 1.985 & 0.996773(66) & 0.998143(34) & 0.998520(28)\\
     0.753 & 2.068 & 0.996702(88) & 0.998143(28) & 0.998522(26)\\
     0.814 & 2.151 & 0.996658(85) & 0.998138(22) & 0.998524(22)\\
      \hline \hline 
    \end{tabular}
  \end{center}
  \caption{
    The ratios of projected amputated vertex function for the axial currents with different actions on the M1 ensemble.
    The quark mass for both fields is taken to be $am_l=0.004$.}
  \label{tab:lambdaAratiosM1}
\end{table}

\begin{table}
  \begin{center}
    \begin{tabular}{c c c c c c}
      \hline\hline\\[-4mm]
      $(ap)^2$ & $p\,[\mathrm{GeV}]$ & $\frac{\mathcal{P}[\Lambda_\text{A}](1.8,1.8)}{\mathcal{P}[\Lambda_\text{A}](1.6,1.6)}$ & $\frac{\mathcal{P}[\Lambda_\text{A}](1.6,1.8)}{\mathcal{P}[\Lambda_\text{A}](1.6,1.6)}$ & $\frac{\mathcal{P}[\Lambda_\text{A}](1.8,1.8)}{\mathcal{P}[\Lambda_\text{A}](1.6,1.8)}$\\\hline
       0.482 & 1.926 & 0.996779(23) & 0.9982202(85) & 0.998555(11)\\
       0.516 & 1.990 & 0.996744(26) & 0.9982053(99) & 0.998539(12)\\
       0.548 & 2.054 & 0.996728(24) & 0.9981981(91) & 0.998525(97)\\
       0.583 & 2.118 & 0.996716(19) & 0.9981914(85) & 0.9985203(79)\\
       0.619 & 2.183 & 0.996719(19) & 0.998189(10)  & 0.9985242(64)\\
      \hline \hline 
    \end{tabular}
  \end{center}
  \caption{
    The ratios of projected amputated vertex function for the axial currents with different actions on the F1 ensemble.
    The quark mass for both fields is taken to be $am_l=0.002144$. }
  \label{tab:lambdaAratiosF1}
\end{table}
\subsection{Strange-quark mass correction}\label{sec:strange_correction}
\begin{figure}
  \center
  \includegraphics[width=0.48\textwidth]{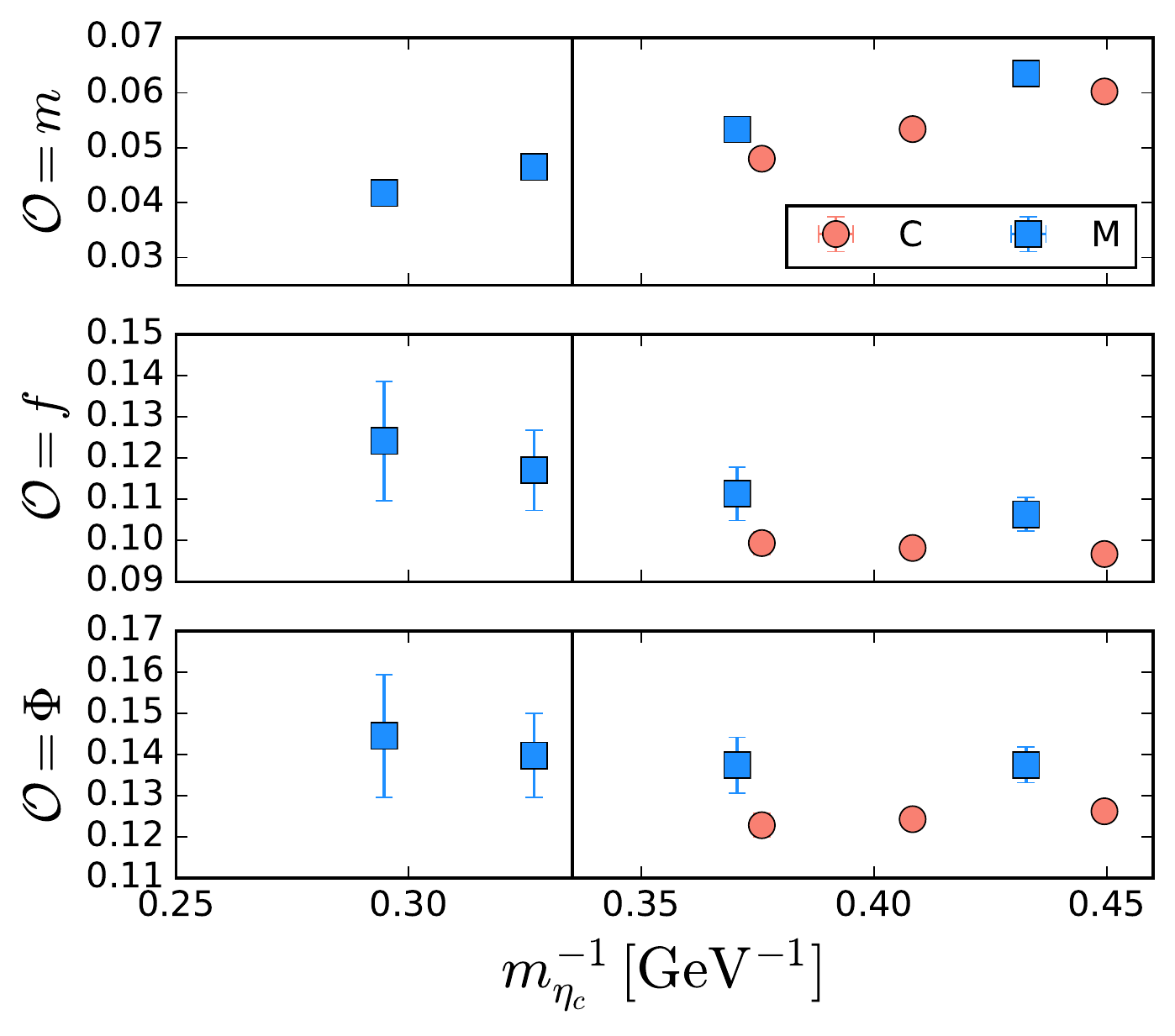}
  \includegraphics[width=0.48\textwidth]{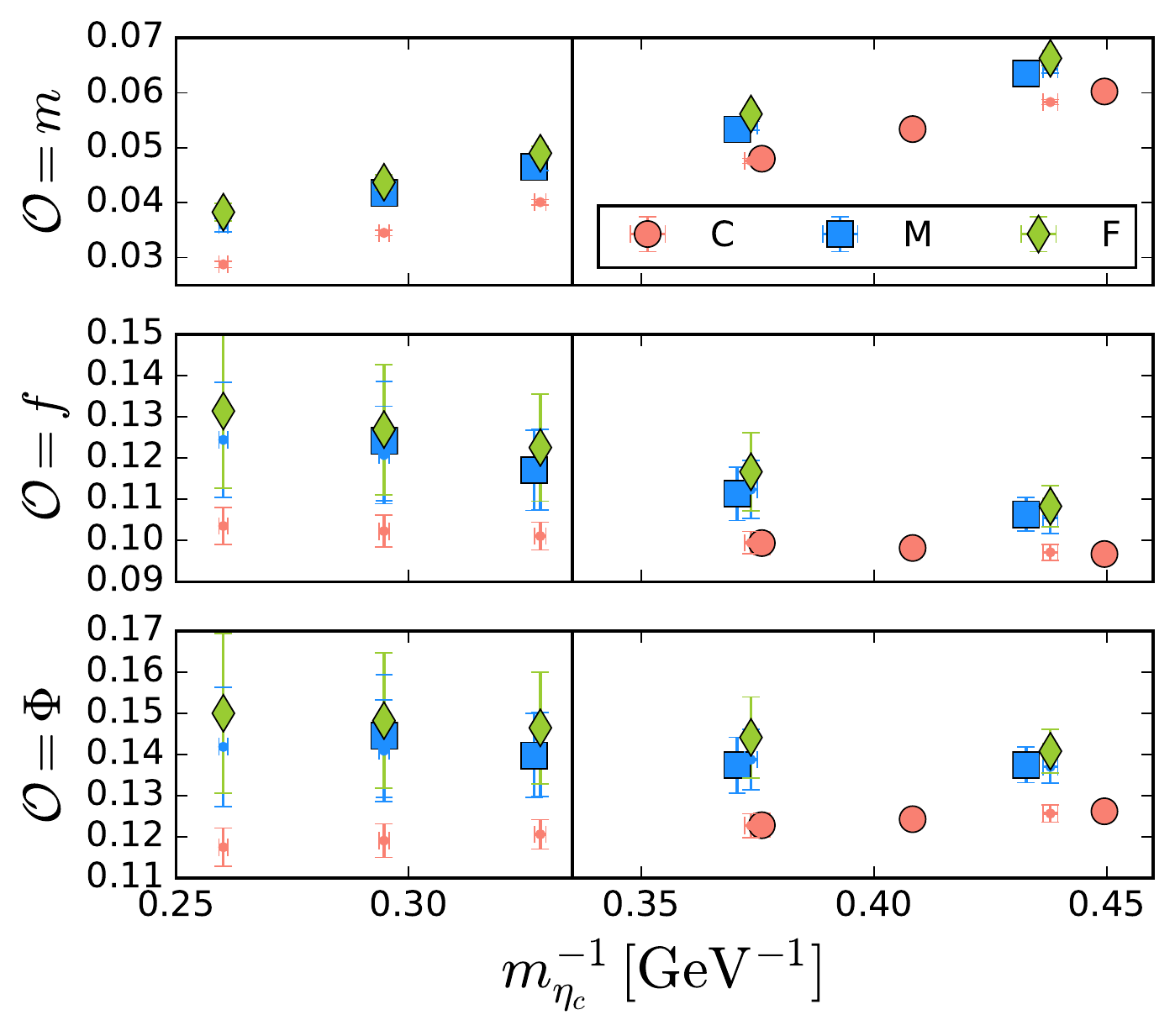}
  \caption{
    \emph{left}: Values of dimensionless parameters $\alpha_\mathcal{O}$ as defined in the text for C1 (red circles) and M1 (blue squares).
    \emph{right}: The same data overlaid with an extrapolation to the heavy quark choices of F1 (small red and blue symbols for C1 and M1, respectively).
    From this, the values of $\alpha_\mathcal{O}$ for the fine ensembles F1 (green diamonds) are found, as outlined in the text.}
  \label{fig:alpha_s}
\end{figure}

We determine the physical strange-quark mass on all ensembles considered here by repeating the global fit to light meson observables detailed in~\cite{RBCUKQCDPhysicalPoint} but including our new fine ensemble F1.
At the time of the data generation the values of $am_s^\mathrm{phys}$ were not yet known for ensembles with near-physical pion masses (C0 and M0) and the new finer ensemble F1. 
To correct for the resulting small mistuning we repeated the simulation of all charmed meson observables on C1 and M1 with both the unitary and the physical valence strange-quark mass.
From this we obtain information on the (small) corrections on C0, M0 and F1.
We define the parameters $\alpha_\mathcal{O}(a,m_h^i)$ for the different observables $\mathcal{O}$ using
\begin{equation}
  \mathcal{O}^\mathrm{phys}(a,m_h) = \mathcal{O}^\mathrm{uni}(a,m_h) \left(1 + \alpha_\mathcal{O}(a,m_h)\, \frac{\Delta m_s}{m_s^\mathrm{phys}} \right),
  \label{eq:definition_alpha}
\end{equation}
where $\Delta m_s \equiv m_s^\mathrm{uni} - m_s^\mathrm{phys}$.
The definition of these $\alpha_\mathcal{O}(a,m_h^i)$ ensures that they are dimensionless and independent of the renormalisation constants.
From the partially quenched data points on C1 and M1 we deduce the values of $\alpha_\mathcal{O}(a,m_h^i)$ for $\mathcal{O} = m_{D_s},\,f_{D_s},\,{f\sqrt{m}}_{D_s}$ for each choice of the simulated heavy quark mass, as shown in the left panel of figure \ref{fig:alpha_s}.

To obtain the values for F1, we need to extrapolate $\alpha_\mc{O}(a,m_h)$ measured on C1 and M1 to $(a,m_h)$ appropriate for F1.
Given the linear behaviour in the inverse heavy quark mass evident from the left panel of figure \ref{fig:alpha_s}, we linearly extrapolate the values for each given lattice spacing to the corresponding $m_h^i(F1)$.
This is then extrapolated to the lattice spacing of F1 by fitting the data to
\eq{
  \alpha_{\mc{O}}(a,m_h^i) = \alpha_{\mc{O}}(0,m_h^i) + C_\alpha a^2\,.
}
The results for this are shown by the green diamonds in the right-hand panel of figure \ref{fig:alpha_s} and summarised in table \ref{tab:alphavalues}.
\begin{table}
\center

\begin{tabular}{|l|l|lll|}
\hline
spacing & $am_h$ & $\alpha_m$ & $\alpha_f$ & $\alpha_\Phi$ \\
\hline\hline
\multirow{3}{*}{coarse}& 0.30  & 0.06026(31) & 0.0967(18) & 0.1262(20)\\
& 0.35  & 0.05341(33) & 0.0982(22) & 0.1243(24)\\
& 0.40  & 0.04801(37) & 0.0994(27) & 0.1229(29)\\
\hline
\multirow{4}{*}{medium}& 0.22  & 0.06353(55) & 0.1064(41) & 0.1375(43)\\
& 0.28  & 0.05335(70) & 0.1113(65) & 0.1375(68)\\
& 0.34  & 0.04650(89) & 0.1171(98) & 0.140(10)\\
& 0.40  & 0.0417(11) & 0.124(14) & 0.145(15)\\
\hline
\multirow{5}{*}{fine}& 0.18  & 0.06630(80) & 0.1083(51) & 0.1408(53)\\
& 0.23  & 0.0562(10) & 0.1167(94) & 0.1442(99)\\
& 0.28  & 0.0490(12) & 0.123(13) & 0.147(14)\\
& 0.33  & 0.0437(14) & 0.127(16) & 0.148(16)\\
& 0.40  & 0.0383(16) & 0.131(19) & 0.150(19)\\
\hline
\end{tabular}
\caption{Values of $\alpha$ for the three observables ($\mc{O}=m_{D_s},f_{D_s},\Phi_{D_s}$). Details about how these were determined can be found in the text.}
\label{tab:alphavalues}
\end{table}

The maximum extent of the strange-quark mass mistuning is present on M0 and given by $\Delta m_s /m_s^\mathrm{phys}=0.048$.
The largest correction (the heaviest charm-quark mass point and the observable $\mathcal{O} = {f\sqrt{m}}_{D_s}$ on M0) is less than $1\%$.

\subsection{Fixing the physical charm quark} \label{subsec:charmmass}
We have a number of possible choices for the meson $H$ that fixes the charm-quark mass.
The ones we will consider are $H=D,D_s$ and $\eta_c$.
Each of these has slightly different advantages and disadvantages attached.
The lattice data for the $D$ meson is comparably noisy and has a strong light-quark-mass dependence, making it difficult to disentangle the extrapolation to physical light-quark masses from the interpolation to the physical charm-quark mass.
The $D_s$ is statistically cleaner and depends less on the light sea-quark mass than the $D$, but we need to correct for a mistuning in the valence strange-quark mass as discussed in section~\ref{sec:strange_correction}.
Finally the $\eta_c$ is statistically the cleanest, but it differs by quark-disconnected Wick contractions with respect to the corresponding physical particle listed by the Particle Data Group (PDG)~\cite{PDG}.
However, this is assumed to be a small effect.\footnote{Ref. \cite{Davies:2010ip} estimates an effect of less than 0.2\% for the contributions due to electromagnetic and quark-disconnected distributions to the mass of $\eta_c$.}

We will investigate all three choices and use the spread as an indication of potential systematic errors.
The masses of these mesons, stated by the PDG~\cite{PDG} are
\al{
  m_{D^\pm}   &= 1.8695(4)\,\mathrm{GeV},\\
  m_{D_s^\pm} &= 1.9690(14)\,\mathrm{GeV},\\
  m_{\eta_c} & = 2.9836(6)\,\mathrm{GeV}.
  \label{eq:PDGmasses}
}
\subsection{Global fit} \label{subsec:global}
Our fit ansatz corresponds to a Taylor expansion around the physical value of the relevant meson masses. 
It is given by
\al{
  \mc{O}(a,m_\pi,m_h) = &\,\, \mc{O}(0,m_\pi^\mathrm{phys},m_h^\mathrm{phys})\\
  & + \left[C_{CL}^0 + C_{CL}^1 \, \Delta m_H^{-1} \right] a^2 \\
  & + \left[C_{\chi}^0 + C_{\chi}^1 \, \Delta m_H^{-1} \right] \left(m_\pi^2 - {m_\pi^{2}}^{\mathrm{phys}} \right)  \\
  & + \left[C_h^0 \right] \Delta m_H^{-1},
  \label{eq:globalchiCL}
}
where $\Delta m_H^{-1} = 1/m_H - 1/m_H^\mathrm{phys}$ and $H = D, D_s$ or $\eta_c$.
This means we simultaneously fit the continuum limit dependence (coefficients $C_{CL}$), the pion mass dependence (coefficients $C_\chi$) and heavy quark dependence (coefficients $C_h$) as well as cross terms (coefficients linear in $\Delta 1/m_h$, i.e. $C_\chi^1$ and $C_{CL}^1$) in one global fit.
The coefficients $C^1_{CL}$ and $C^1_\chi$ capture mass dependent continuum limit and pion mass extrapolation terms.
This arises by expanding $C_{CL}(m_h)$ and $C_{\chi}(m_h)$ in powers of $\Delta m_H^{-1}$. 

\begin{figure}
  \begin{center}
    \includegraphics[width=.48\textwidth]{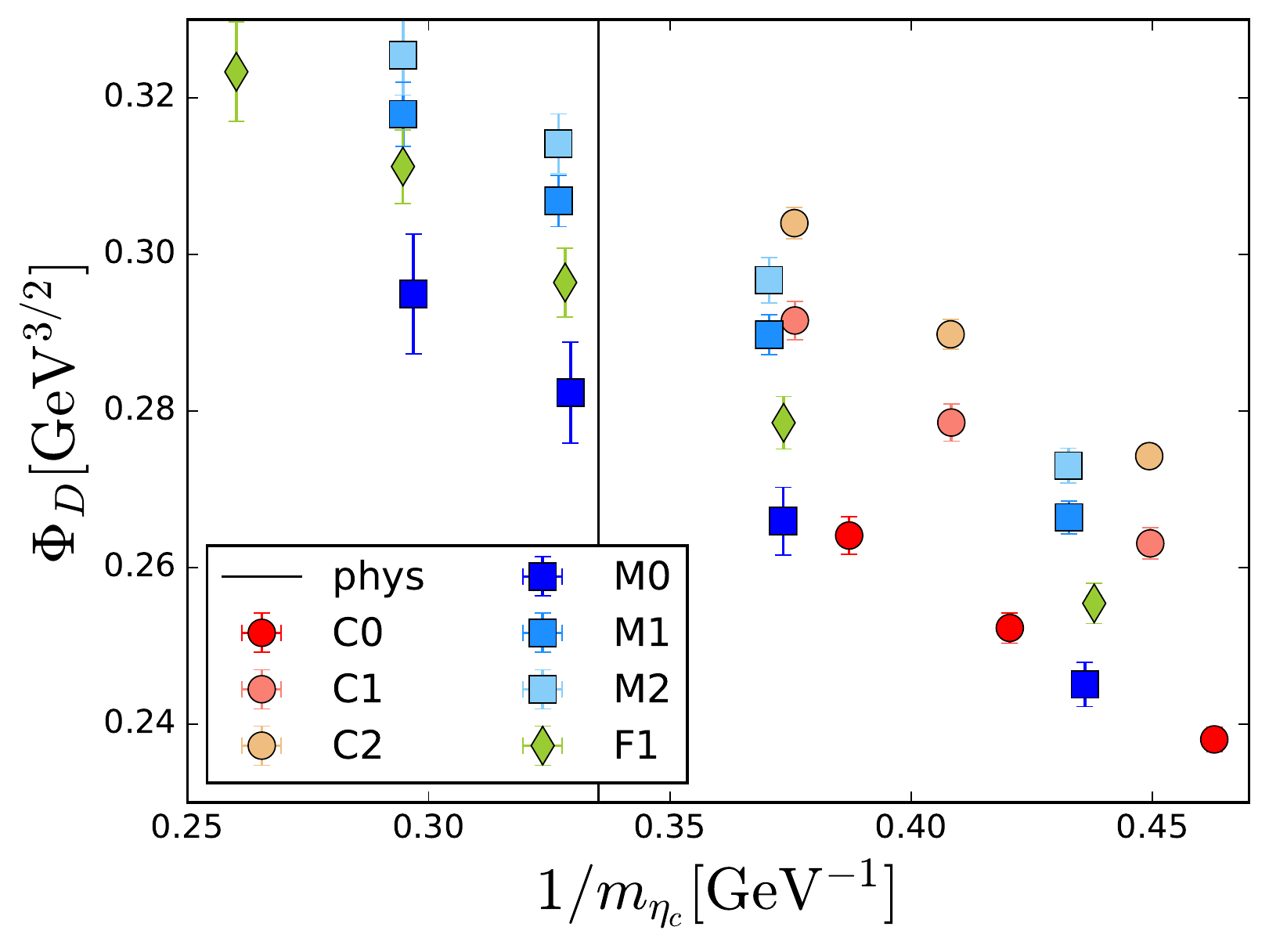}
    \includegraphics[width=.48\textwidth]{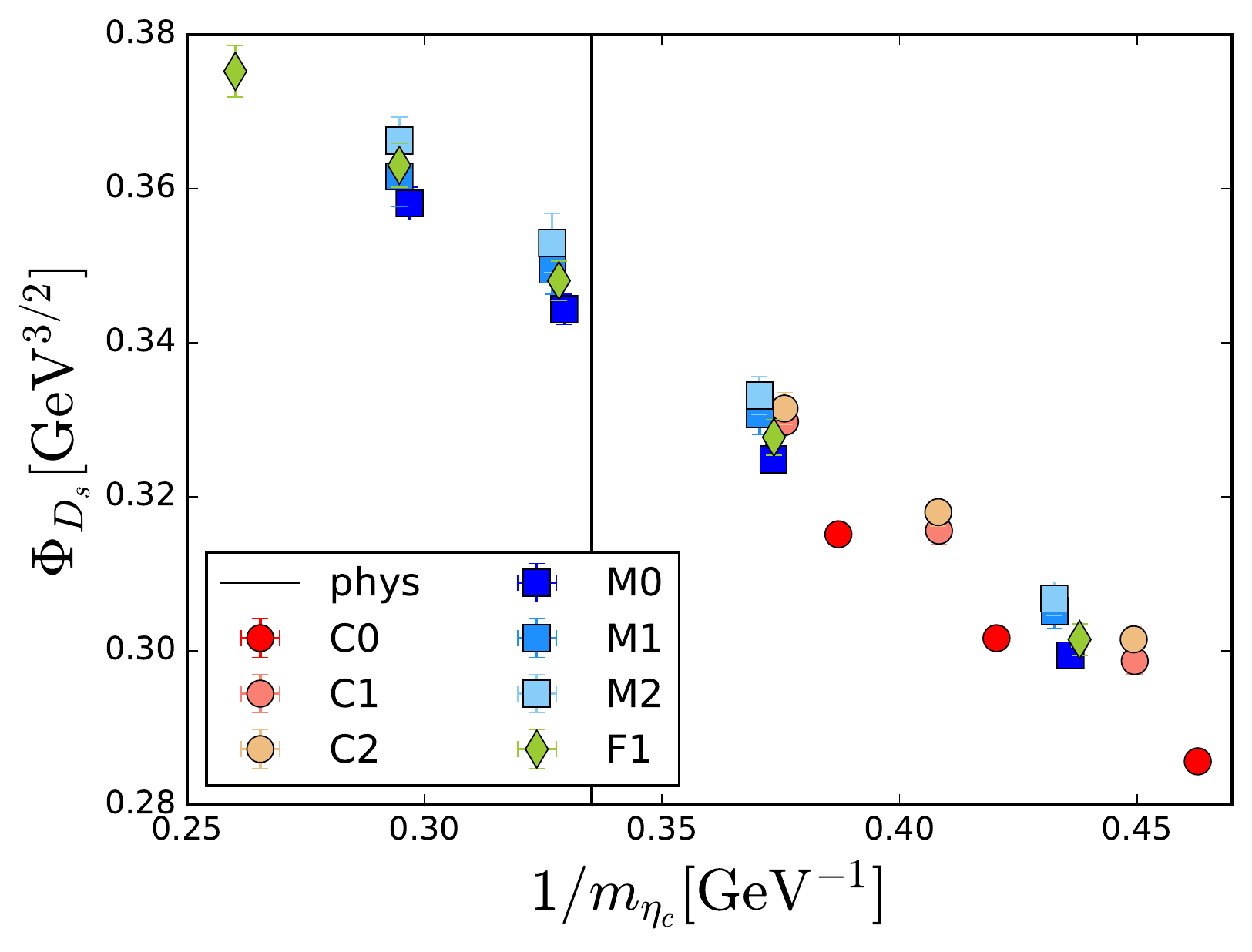}
  \end{center}
  \caption{
    The behaviour of $\Phi_D$ (left) and $\Phi_{D_s}$ (right) as a function of the inverse $\eta_c$ mass for the various ensembles.
    The black vertical line corresponds to the physical value as stated in the PDG~\cite{PDG}.}
  \label{fig:PhiD_PhiDs_data}
\end{figure}

\begin{figure}
  \begin{center}
    \includegraphics[width=.48\textwidth]{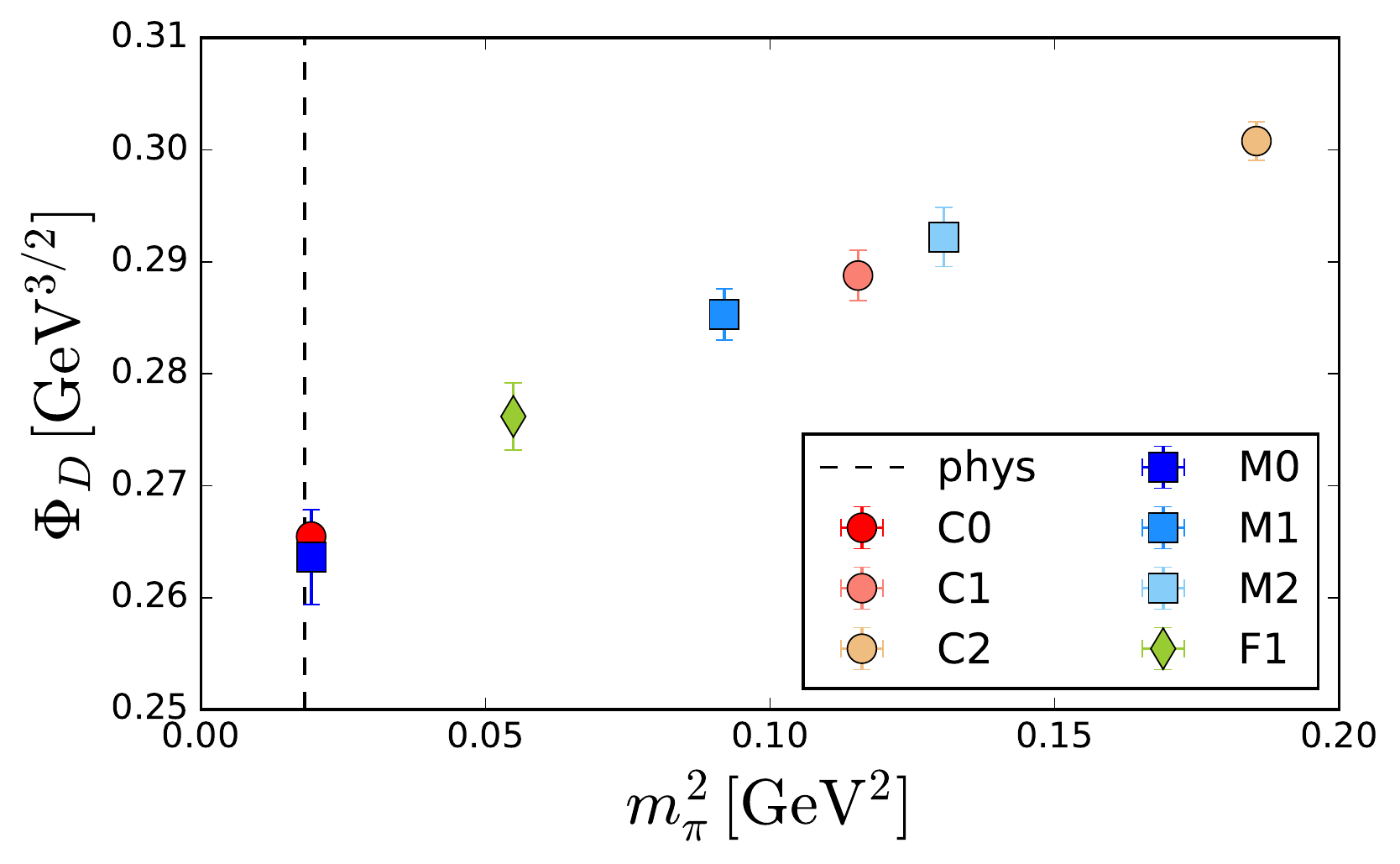}
    \includegraphics[width=.48\textwidth]{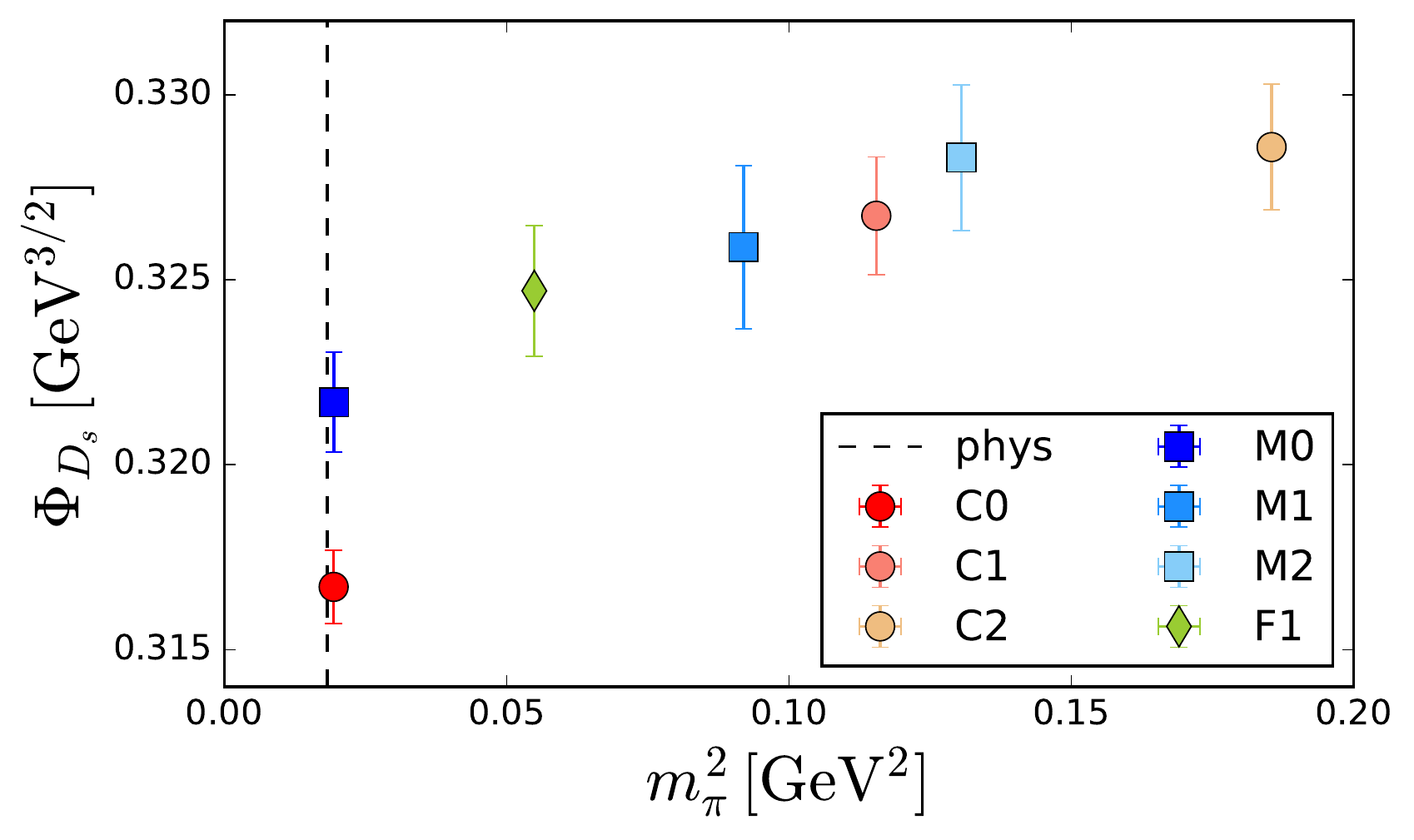}
  \end{center}
  \caption{
    At the current level of statistical resolution the dependence of $\Phi_D$ (left) and $\Phi_{D_s}$ (right) on the light valence- and sea-quark masses is well described by a linear (in $m_\pi^2$) ansatz.
    The plots representatively show data from all ensembles for the reference point $m_{\eta_c} = 2.61\,\mathrm{GeV}$ to which we interpolated linearly in $1/m_{\eta_c}$.}
  \label{fig:ml-dependence}
\end{figure}

When considering the individual decay constants (as opposed to their ratio) we use the quantity $\Phi_{D_{(s)}}=f_{D_{(s)}}\sqrt{m_{D_{(s)}}}$ in the subsequent analysis.
As can be seen in figure~\ref{fig:PhiD_PhiDs_data} this quantity is, within statistical resolution, linear in $1/m_{\eta_c}$.
Also its dependence on the light (sea and valance) quark-mass, as shown exemplarily in figure~\ref{fig:ml-dependence}, is linear irrespective of the heavy-quark mass.

We estimate the systematic uncertainty by limiting the data entering the fit to the case of pion masses not larger than 450, 400 or $350\,\mathrm{MeV}$ in turn.
Another variation we have already mentioned is the choice of the meson that fixed the charm-quark mass.
Finally we can modify the fit form \eqref{eq:globalchiCL} by setting some of the parameters to zero by hand (e.g. $C^1_{CL}$ and $C^1_{\chi}$), which we will do when the data is not sufficiently accurate to resolve them clearly.

\begin{figure}
  \begin{center}
    \includegraphics[width=\textwidth]{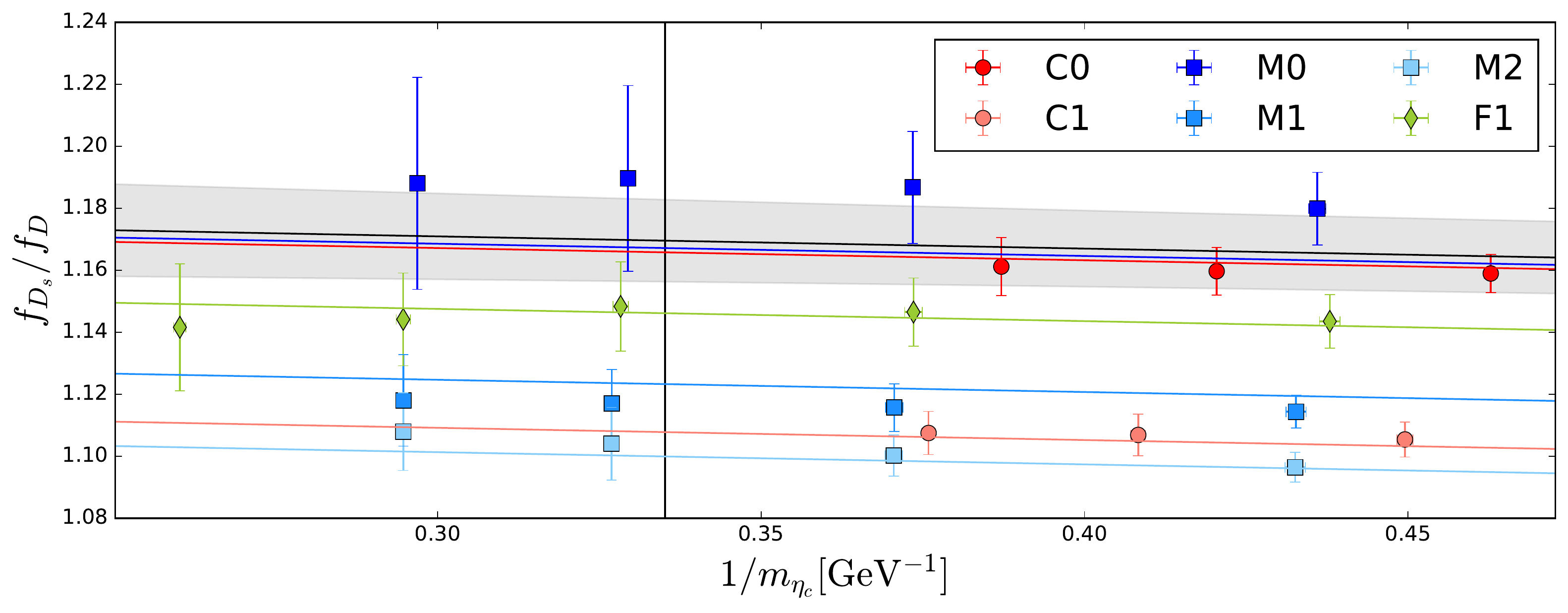}
  \end{center}
  \caption{
    One example for the global fit according to \eqref{eq:globalchiCL} for the case of the observable $f_{D_s}/f_D$.
    In the case presented here the charm-quark mass is fixed by the $\eta_c$ meson and a pion mass cut of $m_\pi<400\mathrm{MeV}$ is employed.
    The grey band shows the fit result at physical pion masses and vanishing lattice spacing.
    The coloured bands correspond to the fit projected to the given pion mass and lattice spacing for the corresponding ensembles.
    In this fit we ignore heavy mass dependent continuum and pion mass terms.}
  \label{fig:global_fDsfD_400MeV_etac_cor}
\end{figure}

\subsubsection{Global fit for ratio of decay constants}
For the ratio of decay constants, fully correlated fits could be achieved.
Figure \ref{fig:global_fDsfD_400MeV_etac_cor} gives one example of such a fully correlated fit for the ratio of decay constants.
The fit shown here has a pion mass cut of $m_\pi < 400 \mathrm{MeV}$ and uses the $\eta_c$ mass to fix the charm-quark mass.
Furthermore, heavy mass dependent coefficients of the continuum limit and the extrapolation to physical pion masses are ignored (i.e. $C_{CL}^1=0$ and $C_\chi^1=0$).

\begin{figure}
  \begin{center}
    \includegraphics[width=\textwidth]{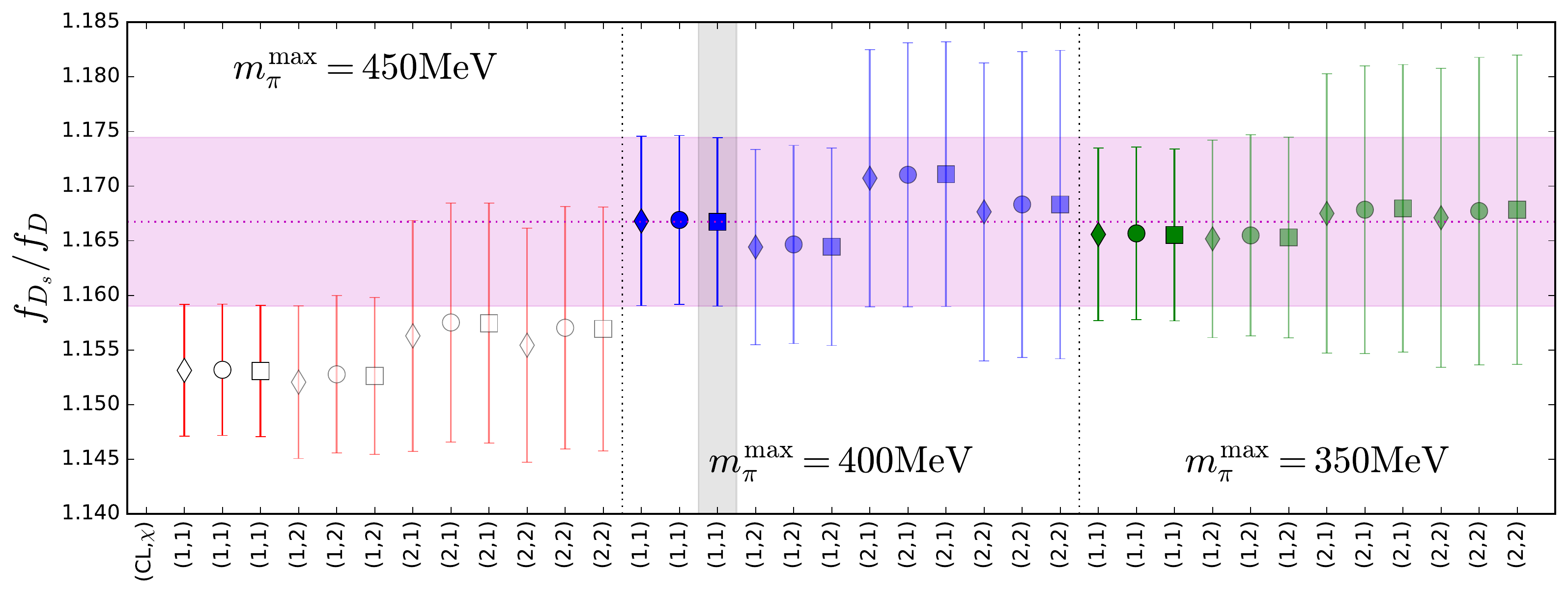}
  \end{center}
  \caption{
    Comparison of the results of the different choices in the global fit.
    The grey and magenta bands highlight the fit shown in figure \ref{fig:global_fDsfD_400MeV_etac_cor}.
    The different symbols indicate different ways of fixing the heavy quark mass, i.e. $H=\,$ $D(\Diamond),\, D_s (\ocircle),$ and (the connected part of) $\eta_c(\square)$.
    Fainter data points indicate that at least one of the heavy mass dependent coefficients is compatible with zero at the one sigma level.
    More detail about the data shown here is given in the text.}
  \label{fig:global_fDsfD_sys}
\end{figure}

Table \ref{tab:global_fits_fDsfD} in appendix~\ref{app:GlobalFitResults} summarises the results of all fit variations for $f_{D_s}/f_D$.
The results of these are also shown in figure \ref{fig:global_fDsfD_sys}.
The red (blue, green) data points correspond to pion mass cuts of $m_\pi^\mathrm{max}=450\,\mathrm{MeV}$ ($400\,\mathrm{MeV}$, $350\,\mathrm{MeV}$).
The different symbols indicate different ways of fixing the heavy quark mass, i.e. $H=\,$ $D(\Diamond),\, D_s (\ocircle),$ and (the connected part of) $\eta_c(\square)$.
Finally the label at the x-axis describes which fit was used by stating the number of coefficients for the continuum limit ($CL$) and pion mass limit ($\chi$) respectively.
E.g. fits results labelled $(2,2)$ correspond to the fit form \eqref{eq:globalchiCL} whilst $(2,1)$ corresponds to keeping two coefficients for the continuum limit extrapolation, but only one coefficient for the pion mass extrapolation by setting $C_{\chi}^1$ to zero.
Cases where one of the coefficients $C^1_{CL}$ and $C^1_{\chi}$ is compatible with zero at the one sigma level are indicated by the corresponding data point being partially transparent.

From the results shown in figure \ref{fig:global_fDsfD_sys} we can make a few observations.
We  find that the ratio of decay constants is insensitive to the way we fix the charm-quark mass.
This is not surprising as the ratio of decay constants does not strongly depend on the heavy quark mass (compare figure \ref{fig:global_fDsfD_400MeV_etac_cor}).
We find that a dependence is observed when including pions with $m_\pi>400\,\mathrm{MeV}$, for this reason we restrict ourselves to $m_\pi\leq 400\, \mathrm{MeV}$.
We can also see that when allowing for heavy mass dependent pion mass and continuum extrapolation terms, these can not be resolved with the present data.
They also do not significantly change the central value of the fit result but increase the statistical error.
This is again not surprising, given the mild behaviour with heavy quark mass displayed by the data.

From this discussion we choose the highlighted fit (i.e. the one presented in figure \ref{fig:global_fDsfD_400MeV_etac_cor}) as our final fit result and as statistical error.
We assign the systematic error arising from the fit form from the maximal spread in the central value of the fit results as we vary the parameters of the fit, maintaining $m_\pi\leq 400 \,\mathrm{MeV}$. 
More precisely, we take the maximal difference between the central value of the preferred fit and the central values of all fits with $m_\pi \leq 400\,\mathrm{MeV}$ displayed in figure \ref{fig:global_fDsfD_sys}.
From this we quote
\eq{
  \frac{f_{D_s}}{f_D} = 1.1667(77)(_{-23}^{+44})_\mathrm{fit},
}
where the first error is statistic and the second error captures the systematic error associated with the chiral-continuum limit as well as the way the charm-quark mass is fixed.

\subsubsection{Global fit for $\Phi_D$ and $\Phi_{D_s}$}
\begin{figure}
  \begin{center}
    \includegraphics[width=\textwidth]{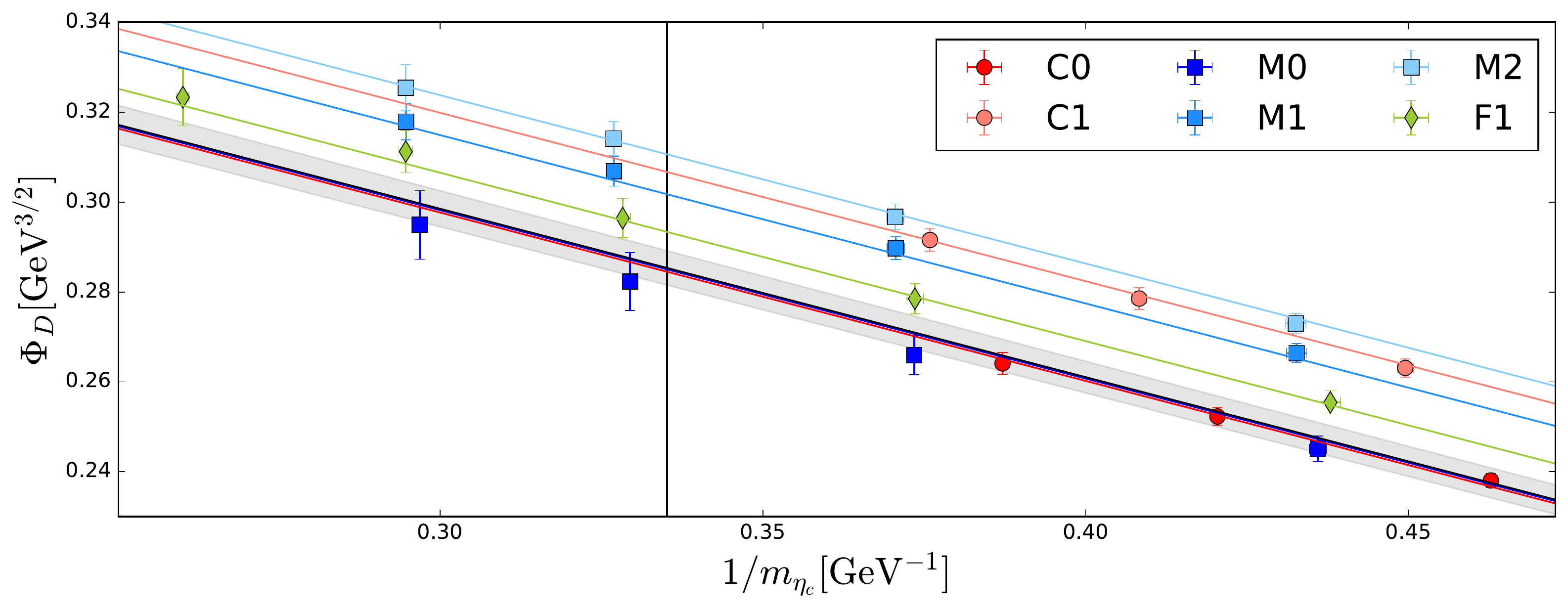}
    \includegraphics[width=\textwidth]{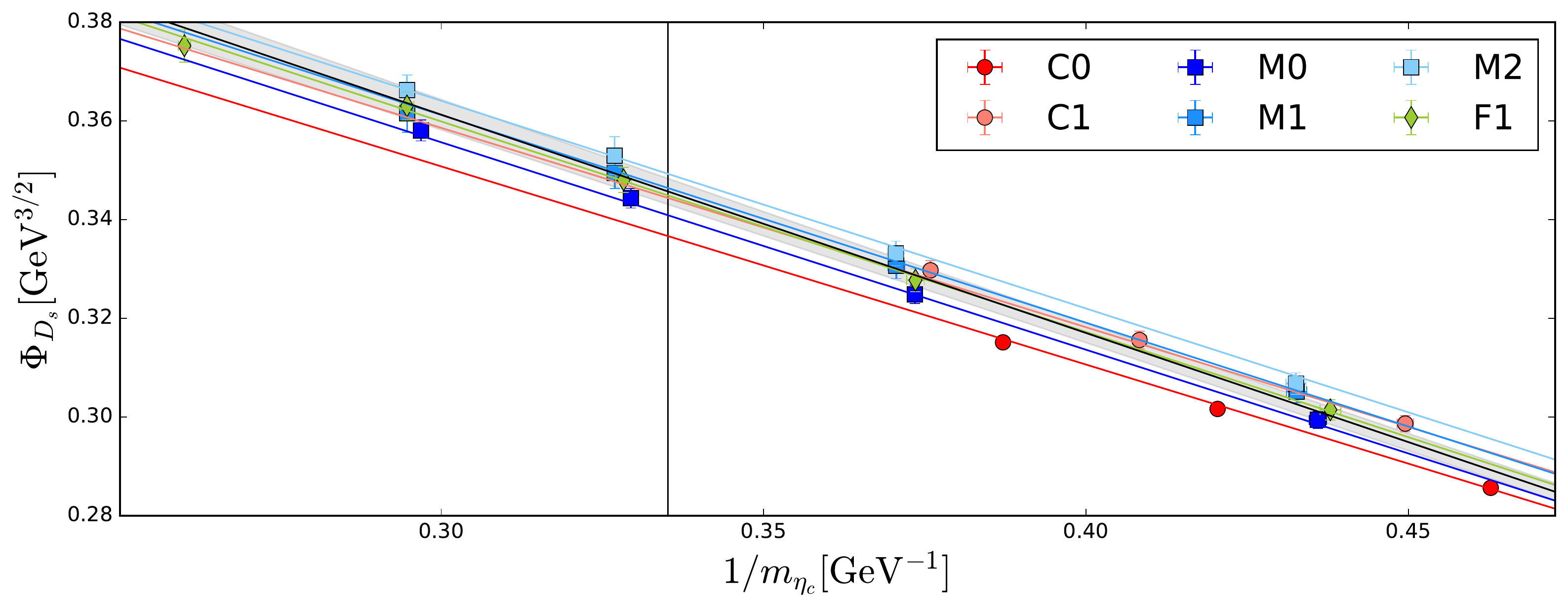}\\
  \end{center}
  \caption{
    Examples for the global fit according to \eqref{eq:globalchiCL} for the case of the observables $\Phi_{D}$ (top) and $\Phi_{D_s}$ (bottom).
    In both cases the charm-quark mass is fixed by the $\eta_c$ meson and a pion mass cut of $m_\pi<400\mathrm{MeV}$ is employed.
    Again, the grey band shows the fit result at physical pion masses and vanishing lattice spacing.
    The coloured bands correspond to the fit projected to the given pion mass and lattice spacing for the corresponding ensembles.
    More details about these fits can be found in the text.}
  \label{fig:global_fmD_fmDs_400MeV_etac_un}
\end{figure}

Figure \ref{fig:global_fmD_fmDs_400MeV_etac_un} shows the chosen fit results for $\Phi_D$ (top) and $\Phi_{D_s}$ (bottom) respectively.
In both cases the heavy quark mass is fixed by the $\eta_c$ mass and a pion  mass cut of $m_\pi\leq400\,\mathrm{MeV}$ is used. Contrary to the fit of the ratio of decay constants, correlated fits of $\Phi_{D}$ and $\Phi_{D_s}$ proved to be unstable. 
Uncorrelated fits were used instead which lead to slightly larger errors. 

\begin{figure}
  \begin{center}
    \includegraphics[width=\textwidth]{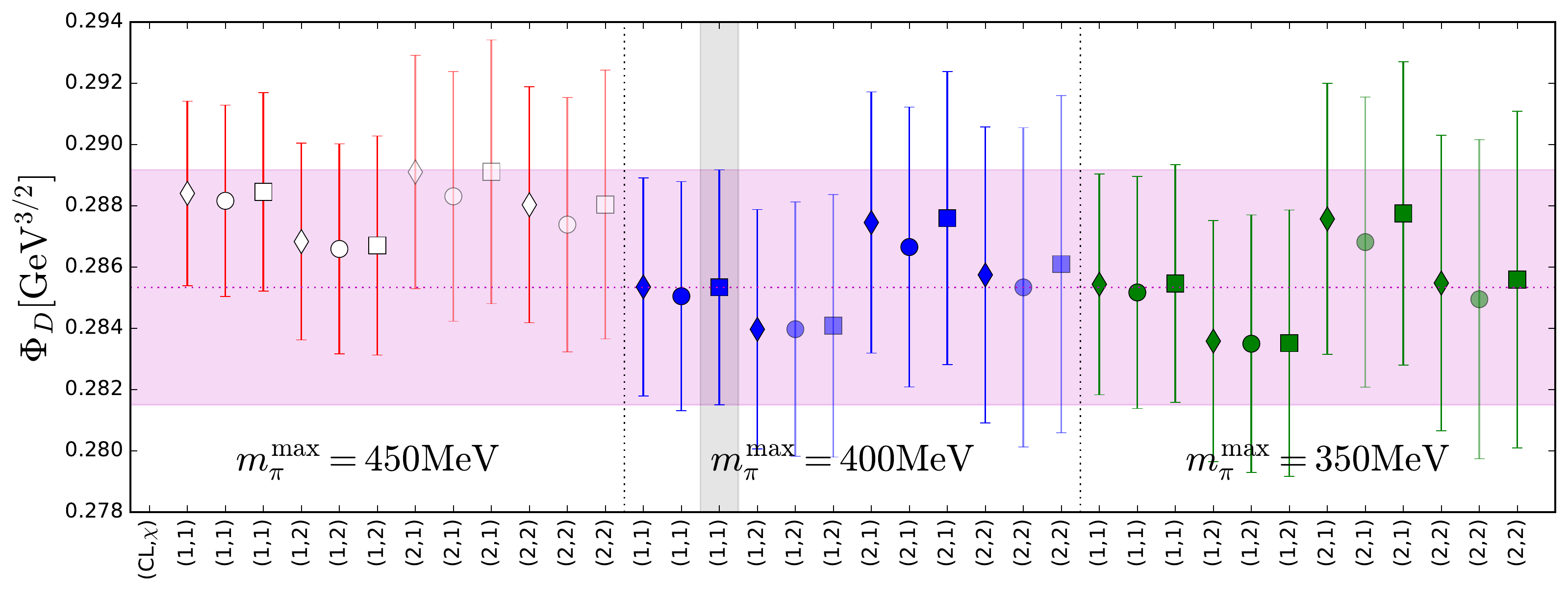}
    \includegraphics[width=\textwidth]{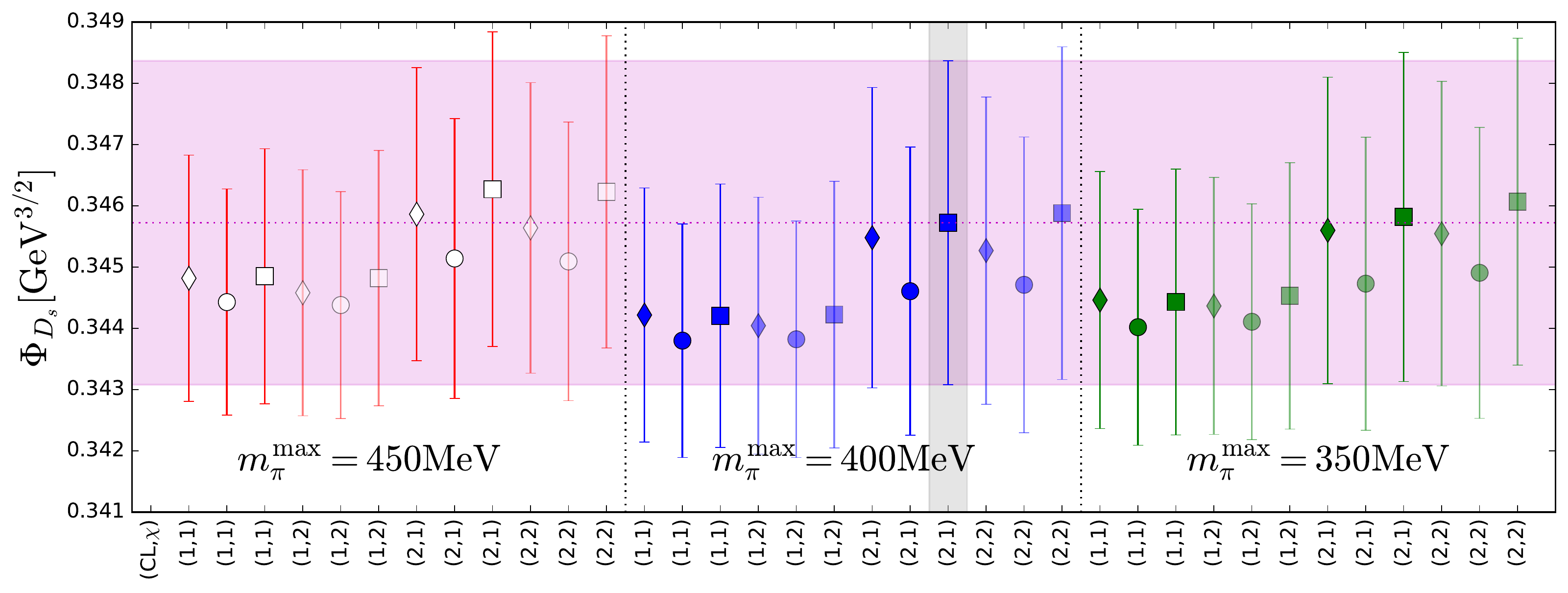}
  \end{center}
  \caption{
    Comparison of the results of the different choices in the global fit for $\Phi_D$ (top) and $\Phi_{D_s}$ (bottom).
    The grey and magenta bands highlight the fit shown in figure \ref{fig:global_fmD_fmDs_400MeV_etac_un}.
    The different symbols indicate different ways of fixing the heavy quark mass, i.e. $H=\,$ $D(\Diamond),\, D_s (\ocircle),$ and (the connected part of) $\eta_c(\square)$.
    Fainter data points indicate that at least one of the heavy mass dependent coefficients is compatible with zero at the one sigma level.
    More detail about the data shown here is given in the text.}
  \label{fig:global_fmD_fmDs_sys}
\end{figure}

Tables \ref{tab:global_fits_fmD} and \ref{tab:global_fits_fmDs} in appendix~\ref{app:nprsmom} summarise the results of all fit variations for $\Phi_D$ and $\Phi_{D_s}$ respectively.
Similar to the previous section we vary the fit parameters to determine the stability of the results.
We find that we can consistently resolve the $C^1_{CL}$ coefficient in the case of $\Phi_{D_s}$, whilst this is less clear in the case of $\Phi_D$.
For this reason we choose $(CL,\chi) = (1,1)$ for the case of $\Phi_D$ and $(CL,\chi) = (2,1)$ for the case of $\Phi_{D_s}$ (see figure~\ref{fig:global_fmD_fmDs_sys}).
Again, little dependence is observed in the case of $m_\pi\leq 400\,\mathrm{MeV}$ so this pion mass cut is used.
The dependence is larger in the case of $\Phi_D$ than for $\Phi_{D_s}$ in agreement with intuition.
We see little dependence in the way the heavy quark mass is fixed, even though (contrary to the ratio of decay constants) the heavy mass dependence is now significant.
Overall we see more variation in the results of the fit than we have for the ratio of decay constants.
Following the same procedure to determine the systematic error associated with the fit as above we find
\al{
  \Phi_D &= 0.2853(38)(_{-18}^{+24})_\mathrm{fit}\,\mathrm{GeV}^{3/2},\\
  \Phi_{D_s} &= 0.3457(26)(_{-19}^{+\hphantom{0}3})_\mathrm{fit}\,\mathrm{GeV}^{3/2}.
}
\subsection{Systematic error analysis} \label{subsec:systematics}
So far we have discussed central values, statistical errors and the systematic errors due to the fit for the value of $\Phi_D$, $\Phi_{D_s}$ and $f_{D_s}/f_D$ and the non-perturbative renormalisation.
We now discuss the remaining systematic error budget due to scale setting, mistuning of the strange-quark mass, finite volume and isospin breaking and continuum limit.

The uncertainty in determining the lattice spacing has been propagated throughout the entire analysis by creating a bootstrap distribution with the width of the error quoted in table \ref{tab:ensembles}.
 The uncertainty in the physical strange-quark masses arising from ref. \cite{RBCUKQCDPhysicalPoint} has been treated in the same way. Both of these are therefore already included in the statistical error.

We have already discussed the systematic error arising from the correction of the mistuning of the strange-quark mass in section \ref{sec:strange_correction} and came to the conclusion that this yields an uncertainty of $9\times 10^{-6}\,\mathrm{GeV}^{3/2}$ for $\Phi_{D_s}$ and $3\times 10^{-4}$ for $f_{D_s}/f_D$.

We estimate the finite size effects by comparing our values of $m_\pi L$ to a study the MILC collaboration has undertaken~\cite{Bazavov:2014wgs}.
They studied the volume dependence of charmed meson decay constants on ensembles with different volumes whilst keeping the lattice spacings and quark masses constant.
This was done for a lattice spacing of $0.12\,\mathrm{fm}$ and pion masses of just above $200\,\mathrm{MeV}$. 
The considered volumes are $2.88\,\mathrm{fm}$, $3.84\,\mathrm{fm}$ and $5.76\,\mathrm{fm}$ corresponding to values of $m_\pi L$ of $3.2$, $4.3$ and $5.4$, respectively. 
For the masses of the $D$ and $D_s$ meson they observed variations of $\lesssim1\,\mathrm{MeV}$ and $\lesssim0.5\,\mathrm{MeV}$, respectively. 
For the decay constants the variations they found are $<0.3\%$ and $<0.15\%$.
Applied to our data, this leads to estimates of the systematic errors of $\delta \Phi_D \approx 0.001\,\mathrm{GeV}^{3/2}$ and $\delta \Phi_{D_s} \approx 0.0006\,\mathrm{GeV}^{3/2}$.
We expect $f_D$ and $f_{D_s}$ to be similarly affected by finite size effects, and therefore expect cancellations in the ratio.
We conservatively take the larger relative error ($0.3\%$) as an estimate for the ratio, yielding $\delta \frac{f_{D_s}}{f_D} \approx 0.0035$.
Given that the minimum value of $m_\pi L$ for our ensembles is $3.8$, results derived from these numbers are a good conservative estimate.

In our simulations we treat the up and down quark masses as degenerate, which is not the case in nature, and neglect electromagnetic effects.
This affects in particular the masses of the mesons we consider.
In principle these effects cannot be disentangled.
In the determination of the decay constants we neglect electromagnetic effects since they are defined as pure QCD quantities.
However, for the determination of the CKM matrix elements these effects will need to be taken into account~\cite{Bazavov:2014wgs,PDG}. 

We devise a systematic error associated to the way we fix the heavy quark mass by considering how much the fit result for $\Phi_D$ changes when we replace the input mass $m_{D^\pm}=1.86961(09)\,\mathrm{GeV}$ by $m_{D^0}=1.86484(05)\,\mathrm{GeV}$~\cite{PDG}.
We estimate the effect of this shift using the fit result of the coefficient $C_h^0$ for the case of $h=D$ and multiplying these by $\abs{m_{D^0}^{-1} - m_{D^\pm}^{-1}}\sim 0.0014\,\mathrm{GeV}^{-1}$.
From this we find $\delta \Phi_D \sim 0.00037\,\mathrm{GeV}^{3/2}$, $\delta \Phi_{D_s} \sim 0.00044\,\mathrm{GeV}^{3/2}$ and $\delta \frac{f_{D_s}}{f_D} \sim 0.00003$.
For the quantity $f_{D_s}/f_D$ this is negligible.
As a probe for the same effect in the light-quark mass fixing, we consider the effect of choosing $m_{\pi^\pm}$ instead of $m_{\pi^0}$ as input mass, i.e. calculating $C_\chi^0 \left(m_{\pi^\pm}^2- m_{\pi^0}^2\right)$. From this we find $\delta \Phi_D \sim 0.00029\,\mathrm{GeV}^{3/2}$, $\delta \Phi_{D_s} \sim 0.00001\,\mathrm{GeV}^{3/2}$ and $\delta \frac{f_{D_s}}{f_D} \sim -0.00080$.
Adding these two effects in quadrature we obtain the values listed in the column $m_u\neq m_d$ in table \ref{tab:systematicerrorbudget}.

\begin{table}
  \center
  \begin{tabular}{|c||c||cc||cccccc|}
    \hline
    \rotatebox[origin=c]{90}{observable} &
    \rotatebox[origin=c]{90}{central} &
    \rotatebox[origin=c]{90}{stat} &
    \rotatebox[origin=c]{90}{\hphantom{aa}total systematic} &
    \rotatebox[origin=c]{90}{fit sys} &
    \rotatebox[origin=c]{90}{finite volume} & 
    \rotatebox[origin=c]{90}{h.o. CL} & 
    \rotatebox[origin=c]{90}{$m_u \neq m_d$} &
    \rotatebox[origin=c]{90}{renormalisation} &
    \rotatebox[origin=c]{90}{strange quark}\\

    & & \multicolumn{2}{c||}{$\times 10^4$} & \multicolumn{6}{c|}{$\times 10^4$}\\
    \hline\hline
    $\Phi_D\,[\mathrm{GeV}^{3/2}]$     & 0.2853 & 38 & $^{+29}_{-24}$ & $_{-18}^{+24}$ & 10\hphantom{.0} & - & \hphantom{0}4.7 & 11\hphantom{.0} & -\\\hline
    $\Phi_{D_s}\,[\mathrm{GeV}^{3/2}]$  & 0.3457 & 26 & $^{+18}_{-26}$ & $_{-19}^{+\hphantom{0}3}$ & \hphantom{0}6\hphantom{.0} & \hphantom{0}7\hphantom{.0} & \hphantom{0}4.4 & 14\hphantom{.0} & \hphantom{0}0.9 \\\hline
    $f_{D_s}/f_D$ & 1.1667 & 77 & $^{+57}_{-43}$ & $_{-23}^{+44}$ & 35\hphantom{.0} & - & \hphantom{0}8\hphantom{.0} & - & \hphantom{0}3\hphantom{.0}\\\hline
  \end{tabular}
  \caption{Summary of the systematic error budget for the quantities $\Phi_D$, $\Phi_{D_s}$ and the ratio of decay constants. Details of the discussion leading to these results can be found in the text.}
  \label{tab:systematicerrorbudget}
\end{table}

Given that the continuum limit coefficient $C_{CL}^0$ is compatible with zero for the fits chosen for $\Phi_D$ ($C_{CL}^0=-0.003(11)\,\mathrm{GeV}^{7/2}$) and $f_{D_s}/f_D=0.005(25)\,\mathrm{GeV}^{2}$, we neglect higher order $O(a^4)$ effects.
For $\Phi_{D_s}$ we find $C_0^{CL} = -0.027(10)\,\mathrm{GeV}^{7/2}$ (the heavy mass dependent continuum limit term vanishes at the physical charm-quark mass).
To estimate the impact of higher order discretisation effect terms we write
\eq{
  \frac{\delta \Phi_{D_s}}{\Phi_{D_s}} = \frac{1}{\Phi_{D_s}} \left[C^0_{CL}a^2 + D^0_{CL} a^4\right] = \frac{C^0_{CL}a^2 }{\Phi_{D_s}} \left[1 + \frac{D^0_{CL}}{C^0_{CL}}  a^2\right].
}
Substituting the numbers for $C_{CL}^0$ and the coarsest and finest lattice spacings we find $C_{CL}^0 a^2/\Phi_{D_s}\sim 0.026$ and 0.010 respectively.
Assuming $D^0_{CL}/C^0_{CL}=(0.5\,\mathrm{GeV})^2$ (i.e. setting the scale such that discretisation effects grow as $a/\Lambda$ with $\Lambda=500\,\mathrm{MeV}$) we find $D^0_{CL}/C^0_{CL} a^2\sim 0.008$ and $\sim 0.003$.
So the residual discretisation effects are 8\% (3\%) of the leading discretisation effects, yielding at most $0.2\%$ of the absolute value.

\begin{figure}
  \begin{center}
    \includegraphics[width=.48\textwidth]{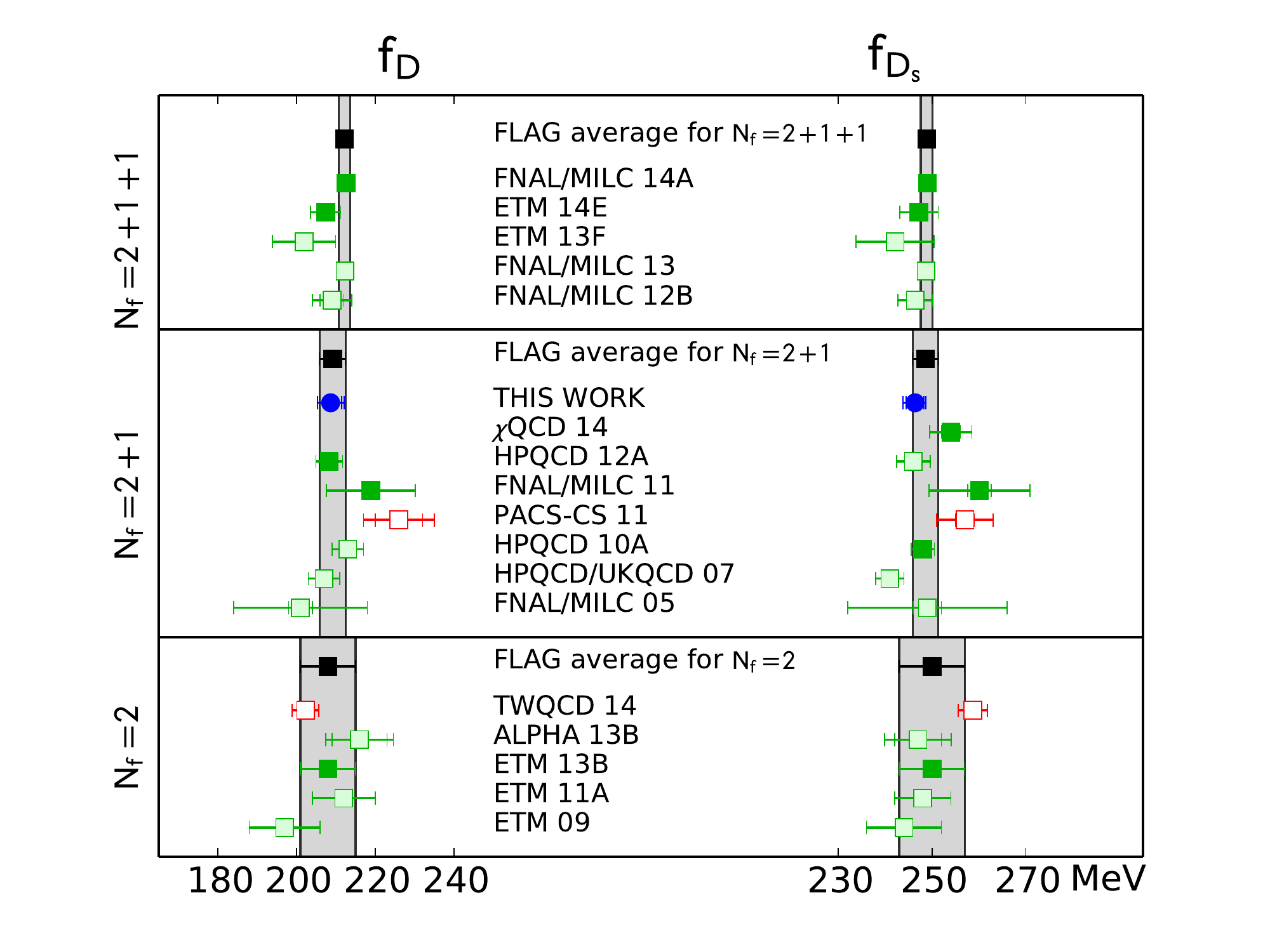}
    \includegraphics[width=.48\textwidth]{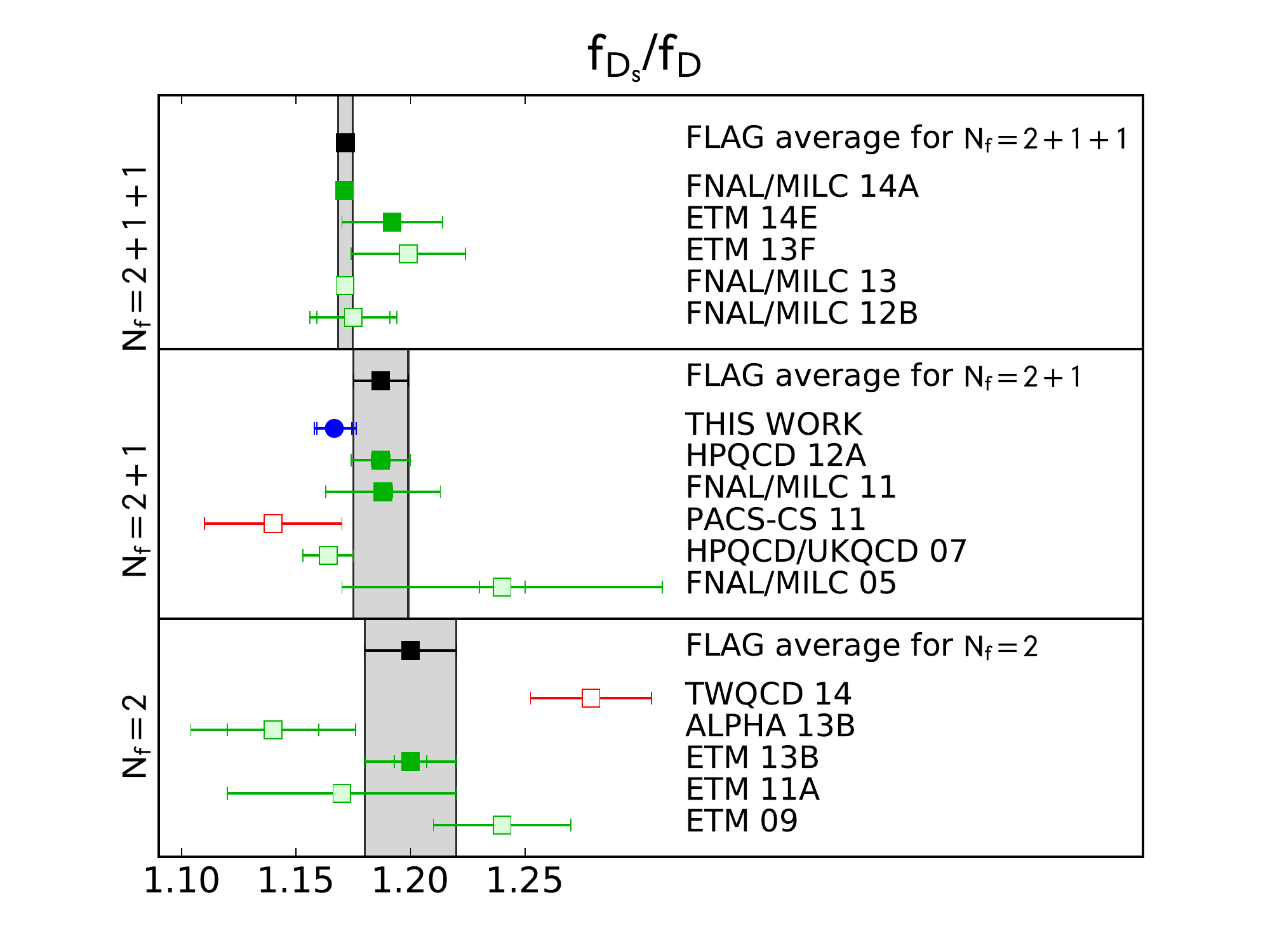}
  \end{center}
  \caption{
    Superposition of our results (blue circles) to the data presented in the most recent FLAG report~\cite{Aoki:2016frl}.
    The small error bar shows the statistic error only, whilst the large error band includes both, the statistic and the systematic error.}
  \label{fig:FLAG_fDfDs_inc_RBCUKQCD}
\end{figure}

Combining these errors in quadrature we arrive at our final values for $\Phi_{D_{(s)}}$.
Using the masses of $D^\pm$ and $D_s^\pm$~\cite{PDG} (compare eq.~\eqref{eq:PDGmasses}) we obtain values for the decay constants $f_{D_{(s)}}$,
\al{
  \Phi_{D} \,= 0.2853(38)_\mathrm{stat}(_{-24}^{+29})_\mathrm{sys}\,\mathrm{GeV}^{3/2} \quad 
  \Rightarrow \quad \,f_{D} &= 208.7(2.8)_\mathrm{stat}(^{+2.1}_{-1.8})_\mathrm{sys}\,\mathrm{MeV}\,,\\
  &\\
  \Phi_{D_s} = 0.3457(26)_\mathrm{stat}(_{-26}^{+18})_\mathrm{sys}\,\mathrm{GeV}^{3/2} \quad
  \Rightarrow \quad f_{D_s} &= 246.4(1.9)_\mathrm{stat}(^{+1.3}_{-1.9})_\mathrm{sys}\,\mathrm{MeV}\,.\\
}
We are now in a position to compare our results to the results found in the literature. 
Adding our results to those presented in the most recent FLAG report~\cite{Aoki:2016frl} we obtain the plots in figure \ref{fig:FLAG_fDfDs_inc_RBCUKQCD}.
The smaller error bar presents the statistic error only, whilst the larger error bar shows the full error (statistic and systematic).
In all cases the error budget is dominated by the statistical error.
We find good agreement with the literature and have errors competitive with the other results displayed in figure \ref{fig:FLAG_fDfDs_inc_RBCUKQCD}
~\cite{
  Bazavov:2014wgs,Carrasco:2014poa,Dimopoulos:2013qfa,Bazavov:2013nfa,Bazavov:2012dg,
  Yang:2014sea,Na:2012iu,Bazavov:2011aa,Namekawa:2011wt,Davies:2010ip,Follana:2007uv,Aubin:2005ar,
  Chen:2014hva,Heitger:2013oaa,Carrasco:2013zta,Dimopoulos:2011gx,Blossier:2009bx
}. 

\section{CKM matrix elements} \label{sec:CKMelements}
\begin{figure}
  \begin{center}
    \includegraphics[width=.48\textwidth]{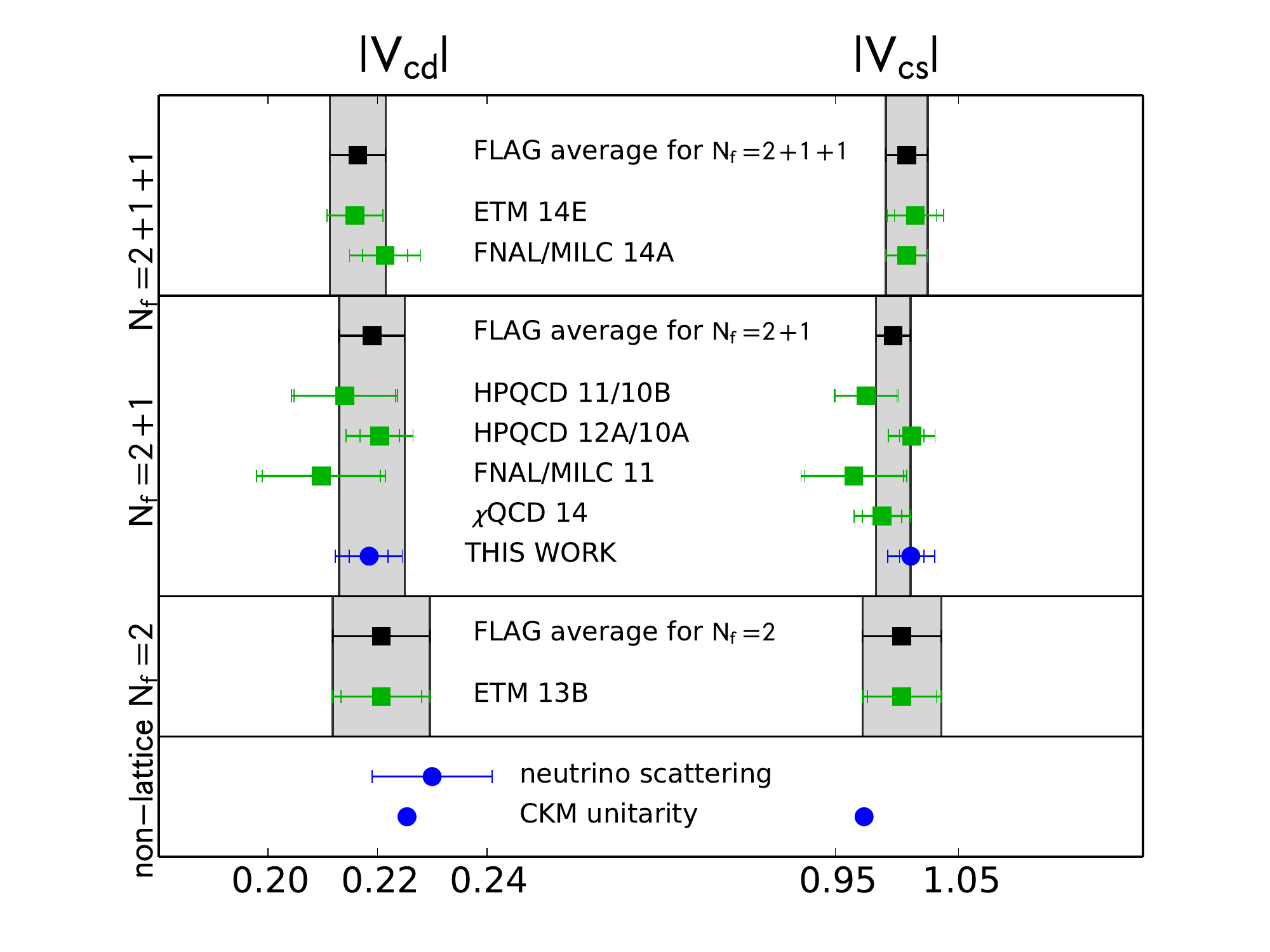}
  \end{center}
  \caption{
    Superposition of our results (blue circles) to the data presented in the most recent FLAG report~\cite{Aoki:2016frl}.
    The smaller error bars of our results show the lattice error only, whilst the large error bands include both, the theoretical and the experimental errors, added in quadrature.}
  \label{fig:FLAG_VcdVcs_inc_RBCUKQCD}
\end{figure}

Having obtained the decay constants, we can make a prediction of the CKM matrix elements $\abs{V_{cd}}$ and $\abs{V_{cs}}$.
However, the values shown in \eqref{eq:expresults} are obtained in nature and therefore we need to adjust these values to those of an isospin symmetric theory.
In other words, the measured decay rate $\abs{V_{cq}}f_{D_q}$ does include electroweak, electromagnetic and isospin breaking effects, so before extracting $\abs{V_{cq}}$ we need to correct the decay rate for these effects.
Ref. \cite{Bazavov:2014wgs} distinguishes between universal long-distance electromagnetic (EM) effects, universal short distance electroweak (EW) effects and structure dependent EM effects.
All of these modify the decay rate to match the experimental value to the theory in which we simulate.
The combined effect of the universal long-distance EM and short-distance EW effects is to lower the decay rate by 0.7\%~\cite{Bazavov:2014wgs,Kinoshita:1959ha,Sirlin:1981ie}.
We adjust the decay rates from \eqref{eq:expresults} and then calculate the CKM matrix elements from this.
We find
\al{
  \abs{V_{cd}} &= 0.2185(50)_\mathrm{exp}(^{+35}_{-37})_\mathrm{lat}, \\
  \abs{V_{cs}} &= 1.011(16)_\mathrm{exp}(^{+11}_{-\hphantom{0}9})_\mathrm{lat}.
}
Again, we can superimpose our results to those obtained in the most recent FLAG report~\cite{Aoki:2016frl}, shown in figure \ref{fig:FLAG_VcdVcs_inc_RBCUKQCD}.
This combines the results of refs. \cite{Bazavov:2014wgs,Carrasco:2014poa,Yang:2014sea,Na:2012iu,Bazavov:2011aa,Davies:2010ip,Carrasco:2013zta,Na:2010uf,Na:2011mc}.
Again we find good agreement between previous works and obtain a competitive error. 
\section{Conclusion and outlook}\label{sec:summary}
In this paper we reported on RBC/UKQCD's first computation of the $D$- and $D_s$-meson decay constants on $N_f=2+1$ domain wall fermion ensembles with physical light quarks and (valence) domain wall charm quarks.
The results for decay constants and CKM matrix elements as summarised in equation \eqref{eq:allresults} derive from a thorough data analysis including in particular a continuum extrapolation over three lattice spacings.
With a precision of 1.6\% ($f_{D}$), 1.0\% ($f_{D_s}$) and 0.7\% ($f_{D_s}/f_{D}$) the results are competitive and establish domain wall fermions as a powerful discretisation for heavy quarks. 
We hope that our results will provide useful input to a wide range of applications in (Beyond) Standard Model phenomenology.

Looking ahead, we are exploring changes in the formulation of the domain wall action, such as gauge link smearing, which we found increases the reach in the heavy quark mass on a given ensemble before cut off effects become substantial~\cite{Boyle:2016lzk}.
This will allow us to do computations directly at the physical charm quark mass also on our coarsest ensemble.

\acknowledgments{
  We would like to thank our RBC and UKQCD collaborators for their support, in particular to Antonin Portelli for interesting and helpful discussions, to Oliver Witzel for valuable comments and careful reading of the manuscript and to Marina Marinkovic for contributions in the early stages of the charm project.

  The research leading to these results has received funding from the European Research Council under the European Union's Seventh Framework Programme (FP7/2007-2013) / ERC Grant agreement 279757 as well as SUPA student prize scheme, Edinburgh Global Research Scholarship and STFC, grants ST/M006530/1, ST/L000458/1, ST/K005790/1, and ST/K005804/1, and the Royal Society, Wolfson Research Merit Awards, grants WM140078 and WM160035.
The authors gratefully acknowledge computing time granted through the STFC funded DiRAC facility (grants ST/K005790/1, ST/K005804/1, ST/K000411/1, ST/H008845/1).
The software used includes the CPS QCD code 
(\verb|http://qcdoc.phys.| \verb|columbia.edu/cps.html|), supported in part by the USDOE SciDAC program;
and the BAGEL
(\verb|http://www2.ph.ed.ac.uk/~paboyle/bagel/Bagel.html|)
assembler kernel generator for high-performance optimised kernels and fermion solvers~\cite{Boyle:2009vp}.

\appendix

\section{Properties of ensemble F1}\label{app:ensemble F1}
Here we present details and properties of ensemble F1 which was generated in order to allow for a continuum limit with three lattice spacings.
It has not appeared in any of RBC/UKQCD's previous analyses.

In table \ref{tab:F1properties} we summarise basic simulation parameters and properties for ensemble F1.
All integrated autocorrelation times were estimated using the technique described in ref. \cite{Wolff:2003sm}.
We have used the same implementation of the exact hybrid Monte Carlo algorithm for the ensemble generation as in ref. \cite{RBCUKQCDPhysicalPoint}, with five intermediate Hasenbusch masses, (0.005,0.017, 0.07 , 0.18, 0.45), for the two-flavor part of the algorithm.
A rational approximation was used for the strange quark determinant.
See ref. \cite{RBCUKQCDPhysicalPoint} for more details.
Figures~\ref{fig:Q} and~\ref{fig:w0 t0} show the Monte Carlo evolution and histograms of the topological charge (measured with the GLU package~\cite{Hudspith_code}) and the Wilson flow scales $w_0$ and $\sqrt{t_0}$, respectively.
The measured autocorrelation time motivates our choice to separate evaluations of observables on ensemble F1 in the main part of this paper by 40 molecular dynamics steps (for comparison: the separation is 40 on C1 and 20 on all remaining ensembles).

\begin{table}
  \centering
  \begin{tabular}{lll}
    \hline\hline\\[-4mm]
    $L^3\times T \times L_s$			&$48^3\times 96\times 12/a^4$\\
    $\beta$					&2.31\\
    $\alpha$ 					&2\\
    steps per HMC traj. 			&10\\
    $\delta\tau$				&$ 0.1$\\
    Metropolis acceptance			&93\%\\ 
    Plaquette expectation value 		&0.6279680(23)		&$\tau_{\rm int}=3.2(5)$\\
    $w_0/a$					&2.3917(25)		&$\tau_{\rm int}=32(9)$\\
    $\sqrt{t_0}/a$				&2.03921(98)		&$\tau_{\rm int}=27(7)$\\
    $am_\pi$					&0.08446(18)\\
    $am_K$					&0.18600(22)\\
    $am_{\rm res}$ ($am_q=am_l$)			&0.0002290(19)\\
    \hline\hline
  \end{tabular}
  \caption{Basic properties of the F1 ensemble.}
  \label{tab:F1properties}
\end{table}

\begin{figure}
  \centering
  \includegraphics[width=7cm]{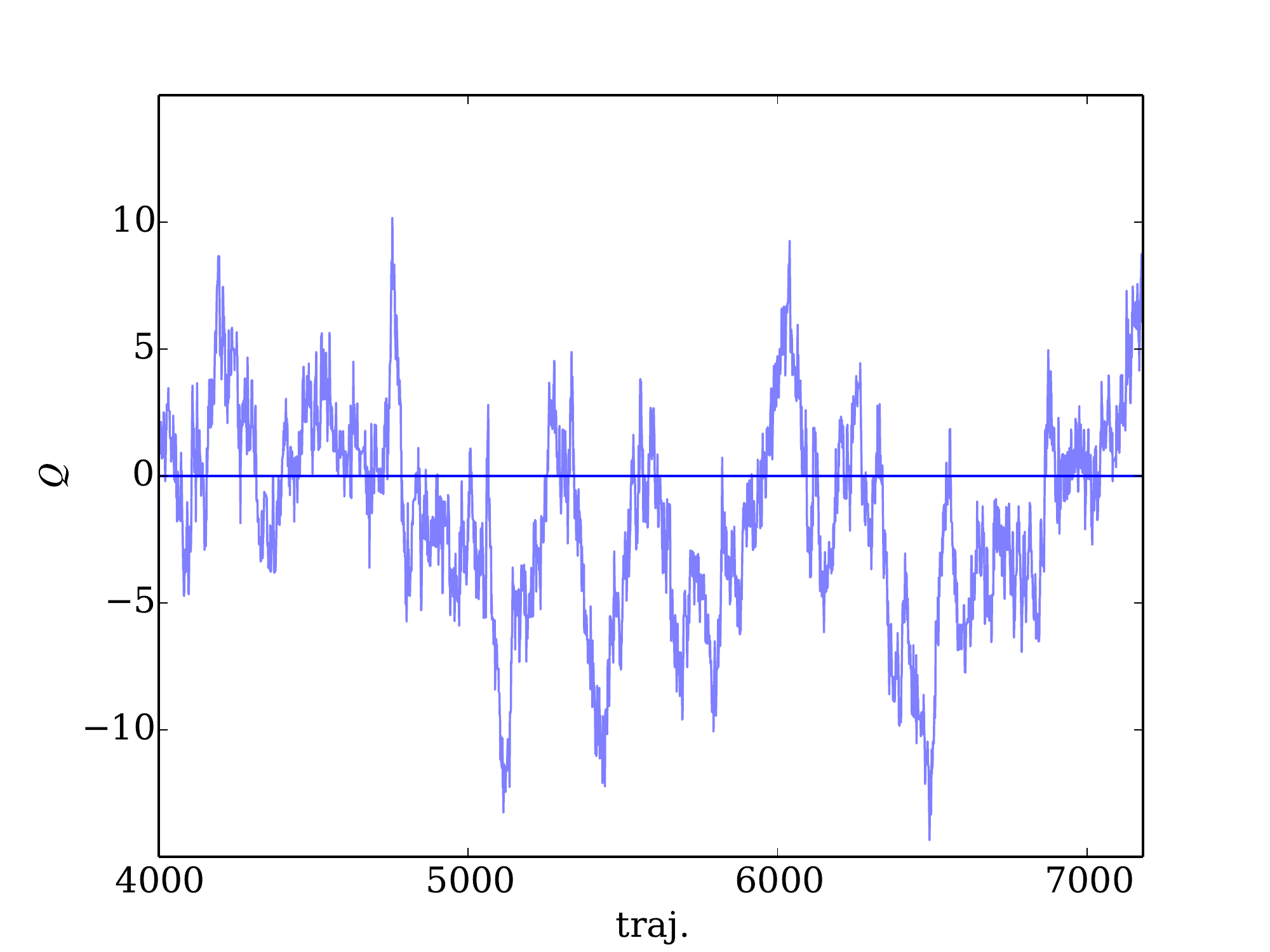}
  \includegraphics[width=7cm]{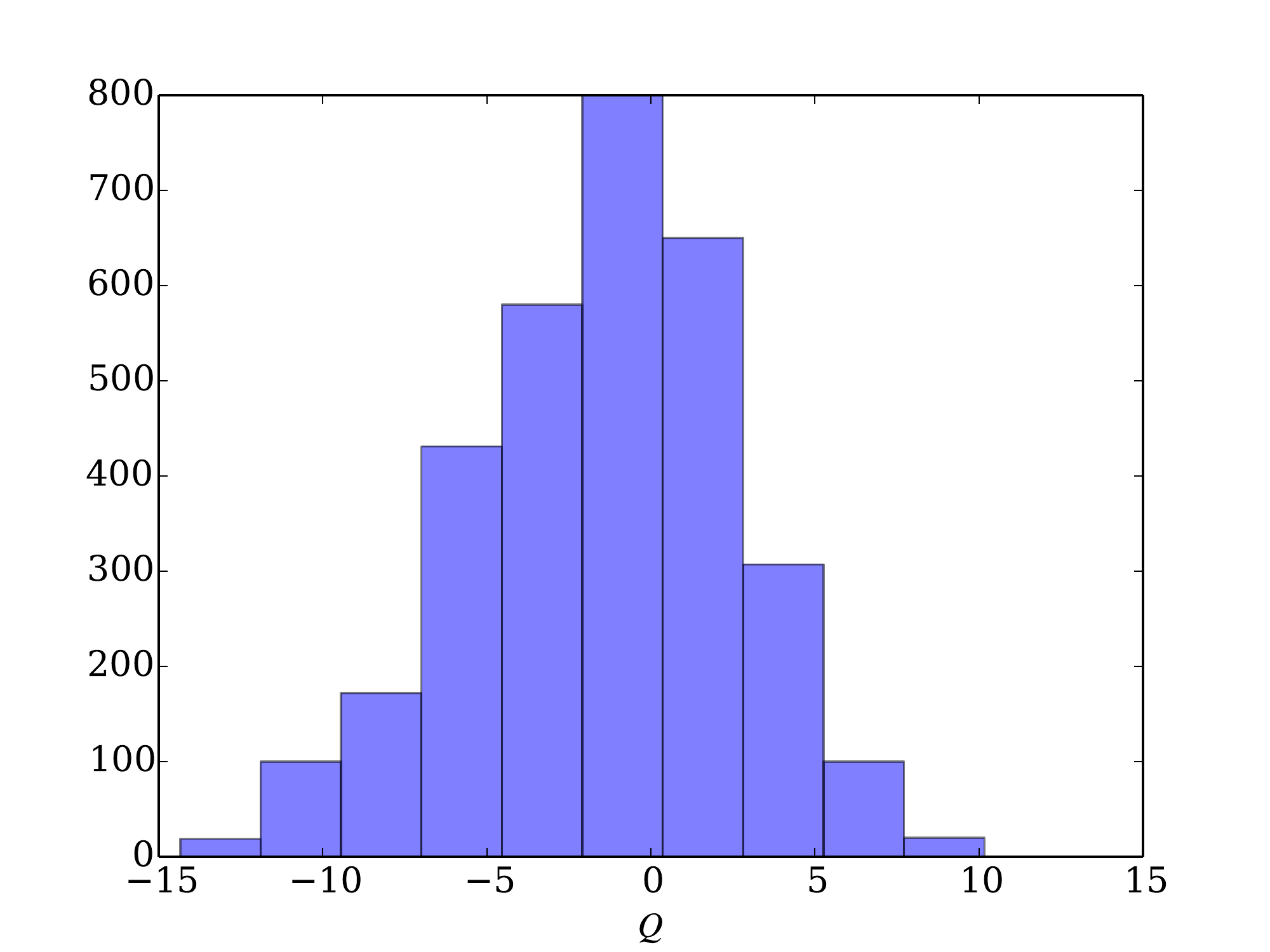}\\
  \caption{
    Monte Carlo history and histogram of the topological charge $Q$.
    The corresponding estimated integrated autocorrelation time for $Q$ is $\tau_{\rm int}=60(2)$ and for $Q^2$ it is $\tau_{\rm int}=34(9)$.}
  \label{fig:Q}
\end{figure}

\begin{figure}
  \centering
  \includegraphics[width=7cm]{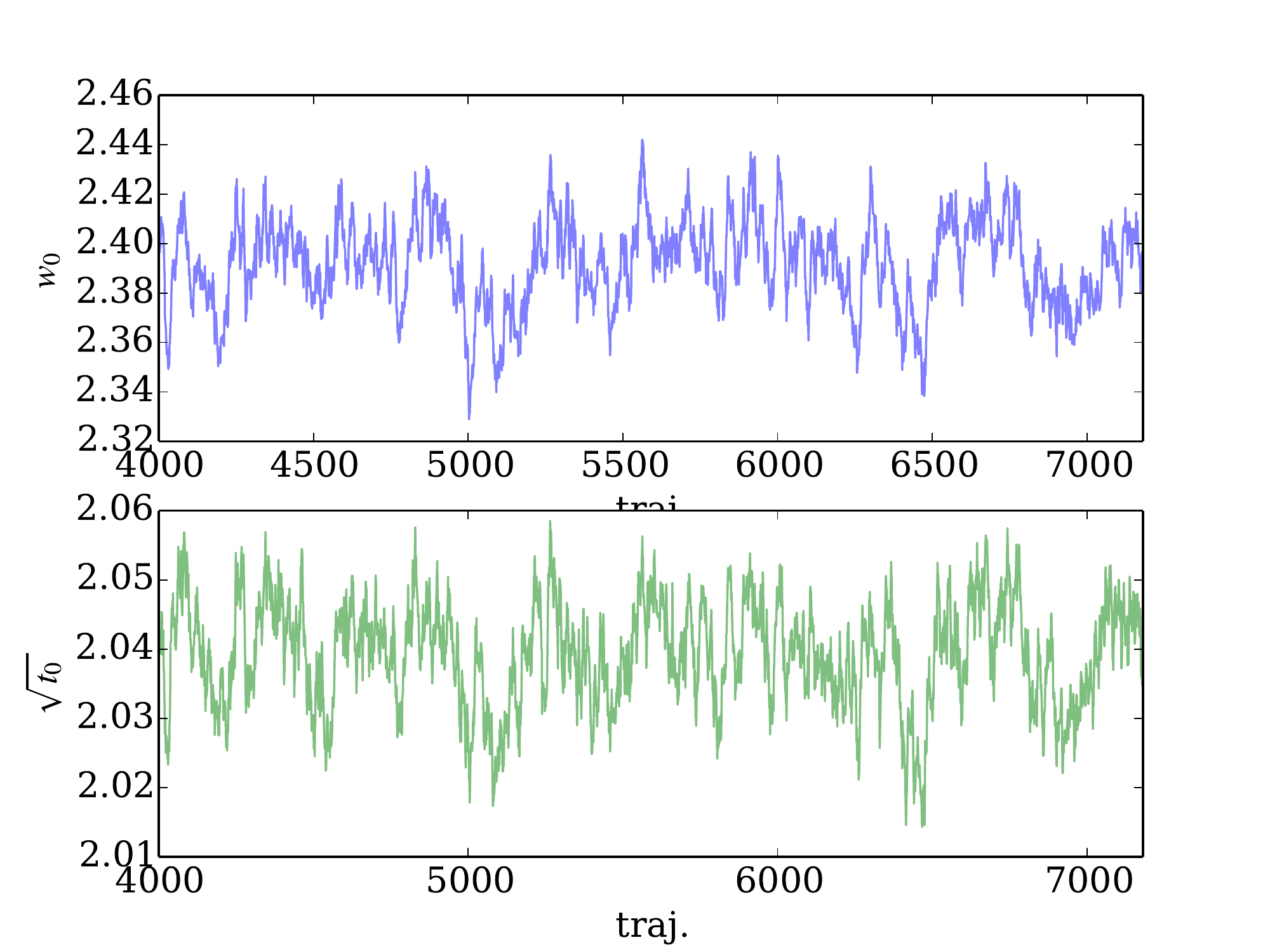}
  \includegraphics[width=7cm]{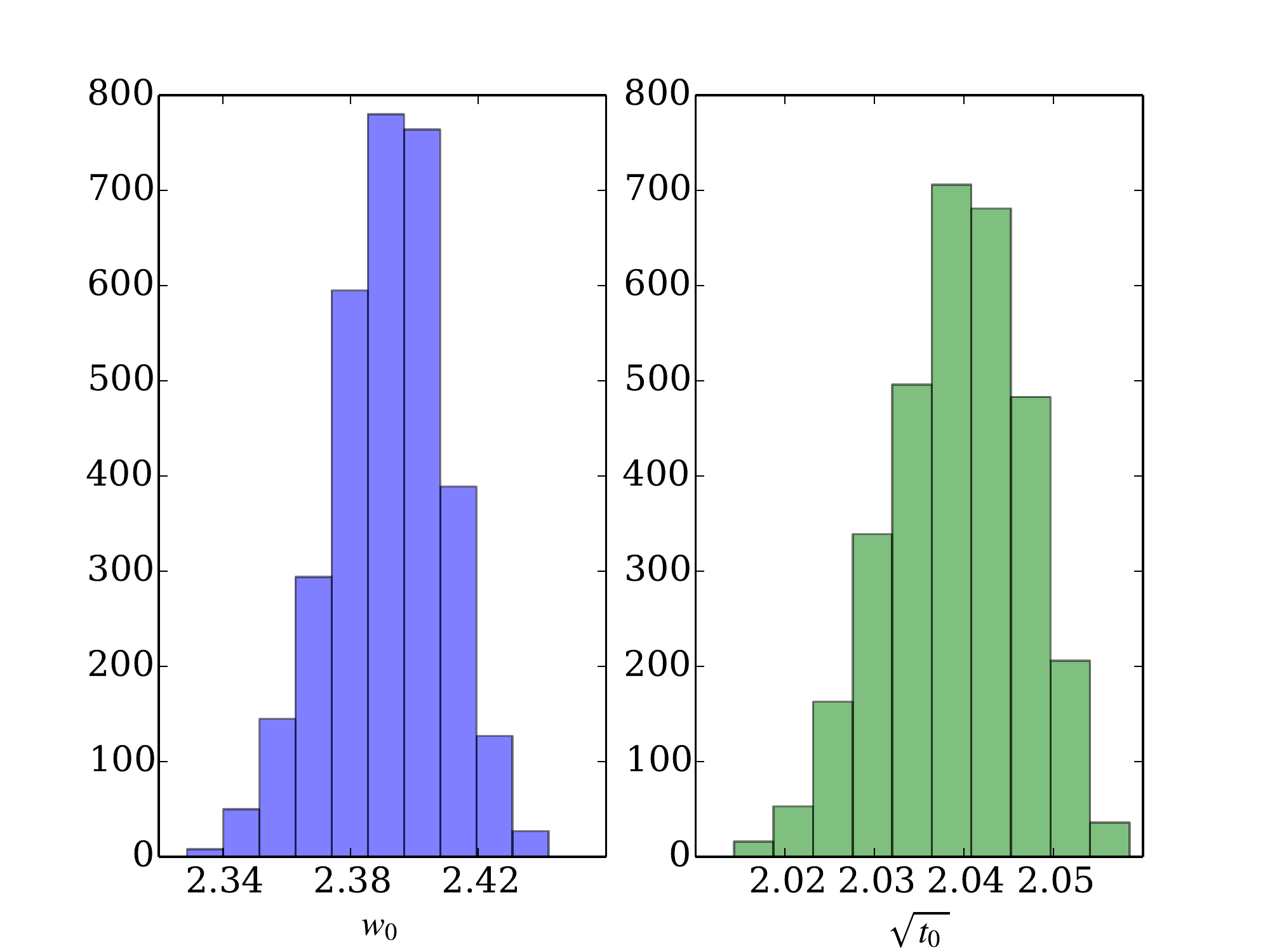}
  \caption{
    \emph{left}: Monte Carlo history of Wilson flow scales $w_0$ and $\sqrt{t_0}$, respectively.
    \emph{right}: The corresponding histograms of $w_0$ and $\sqrt{t_0}$, respectively.}
  \label{fig:w0 t0}
\end{figure}
\section{Correlator fit results}\label{app:correlatorfits}
Tables \ref{tab:D_dynamical}, \ref{tab:Ds_dynamical} and \ref{tab:etac_dynamical} summarise the fit ranges and the fit results of the correlation function fits to the heavy-light, heavy-strange and heavy-heavy pseudoscalar mesons, respectively. Since the different channels may have a slightly different approach to the ground state (compare figure \ref{fig:plateaux}) we quote the smaller value of $t_\mathrm{min}$. In all cases $t_\mathrm{min}^{AP} \leq t_\mathrm{min}^{PP}$. The $t_\mathrm{max}$ quoted is the first time slice which is \emph{not} included in the fit.

\begin{table}
\center
\small
\begin{tabular}{|l|rr||ll|ll|}
\hline
ens & \multicolumn{2}{c||}{$[t_\mathrm{min},t_\mathrm{max})$} & $am_D$ & $af^\mathrm{bare}_D$ & $m_D\,[\mathrm{GeV}]$ & $f^\mathrm{ren}_D\,[\mathrm{GeV}]$ \\
\hline
\multirow{3}{*}{C0} & 8 & 30  & 0.82242(82) & 0.16211(85) & 1.4224(33) & 0.1996(11)\\
 & 8 & 30  & 0.8995(10) & 0.1643(11) & 1.5558(37) & 0.2023(14)\\
 & 8 & 30  & 0.9727(12) & 0.1654(13) & 1.6823(41) & 0.2036(17)\\
\hline
\multirow{3}{*}{C1} & 8 & 25  & 0.83111(95) & 0.1687(10) & 1.4834(44) & 0.2160(14)\\
 & 8 & 25  & 0.9075(11) & 0.1710(12) & 1.6198(49) & 0.2188(16)\\
 & 8 & 22  & 0.9802(11) & 0.1722(12) & 1.7495(52) & 0.2204(16)\\
\hline
\multirow{3}{*}{C2} & 8 & 25  & 0.84129(65) & 0.17468(62) & 1.5015(43) & 0.2238(10)\\
 & 8 & 25  & 0.91722(77) & 0.17679(80) & 1.6370(47) & 0.2265(12)\\
 & 8 & 22  & 0.98995(85) & 0.17851(86) & 1.7669(51) & 0.2287(13)\\
\hline
\hline
\multirow{4}{*}{M0} & 11 & 41  & 0.63066(95) & 0.1146(11) & 1.4875(51) & 0.2010(21)\\
 & 11 & 41  & 0.7259(13) & 0.1159(17) & 1.7122(61) & 0.2032(31)\\
 & 12 & 41  & 0.8146(18) & 0.1162(25) & 1.9214(72) & 0.2037(44)\\
 & 12 & 41  & 0.8975(22) & 0.1156(28) & 2.1168(81) & 0.2027(50)\\
\hline
\multirow{4}{*}{M1} & 9 & 29  & 0.63804(77) & 0.12168(65) & 1.5206(57) & 0.2160(14)\\
 & 9 & 27  & 0.73315(92) & 0.12345(80) & 1.7473(66) & 0.2192(16)\\
 & 9 & 27  & 0.8211(12) & 0.1235(11) & 1.9570(75) & 0.2193(20)\\
 & 9 & 27  & 0.9027(14) & 0.1221(14) & 2.1513(83) & 0.2167(25)\\
\hline
\multirow{4}{*}{M2} & 9 & 32  & 0.64237(80) & 0.12424(73) & 1.5310(57) & 0.2207(15)\\
 & 9 & 32  & 0.73731(98) & 0.12603(97) & 1.7572(66) & 0.2238(19)\\
 & 9 & 32  & 0.8252(12) & 0.1261(13) & 1.9668(76) & 0.2240(24)\\
 & 9 & 32  & 0.9068(16) & 0.1247(17) & 2.1611(85) & 0.2214(32)\\
\hline
\hline
\multirow{5}{*}{F1} & 11 & 42  & 0.53696(70) & 0.09913(79) & 1.4895(57) & 0.2093(18)\\
 & 11 & 41  & 0.61936(88) & 0.1006(10) & 1.7181(67) & 0.2125(23)\\
 & 11 & 41  & 0.6960(11) & 0.1010(13) & 1.9307(77) & 0.2133(29)\\
 & 11 & 37  & 0.7682(12) & 0.1010(13) & 2.1309(85) & 0.2132(29)\\
 & 11 & 37  & 0.8612(16) & 0.0991(18) & 2.3889(98) & 0.2092(38)\\
\hline
\end{tabular}
\caption{Fit results for the masses and decay constants of the $D$ meson on all ensembles. The second set of columns is renormalised where the renormalisation constants are obtained from the light-light conserved current as described in the text.}
\label{tab:D_dynamical}
\end{table}

\begin{table}
\center
\small
\begin{tabular}{|l|rr||ll|ll|}
\hline
ens & \multicolumn{2}{c||}{$[t_\mathrm{min},t_\mathrm{max})$} & $am_{D_s}$ & $af^\mathrm{bare}_{D_s}$ & $m_{D_s}\,[\mathrm{GeV}]$ & $f^\mathrm{ren}_{D_s}\,[\mathrm{GeV}]$ \\
\hline\hline
\multirow{3}{*}{C0} & 12 & 41  & 0.88227(13) & 0.18809(17) & 1.5249(37) & 0.23134(64)\\
 & 12 & 41  & 0.95671(16) & 0.19073(22) & 1.6536(40) & 0.23458(67)\\
 & 12 & 37  & 1.02773(18) & 0.19225(25) & 1.7765(42) & 0.23645(69)\\
\hline
\multirow{3}{*}{C1} & 9 & 32  & 0.87662(43) & 0.18654(45) & 1.5646(49) & 0.23880(98)\\
 & 9 & 32  & 0.95114(46) & 0.18923(53) & 1.6976(53) & 0.2422(11)\\
 & 9 & 32  & 1.02215(51) & 0.19071(62) & 1.8243(56) & 0.2441(11)\\
\hline
\multirow{3}{*}{C2} & 9 & 32  & 0.87804(42) & 0.18799(43) & 1.5671(49) & 0.24085(96)\\
 & 9 & 32  & 0.95235(48) & 0.19040(56) & 1.6998(52) & 0.2439(11)\\
 & 9 & 32  & 1.02309(56) & 0.19146(74) & 1.8260(56) & 0.2453(12)\\
\hline
\hline
\multirow{4}{*}{M0} & 14 & 44  & 0.678132(88) & 0.135920(96) & 1.5946(56) & 0.23711(92)\\
 & 14 & 44  & 0.77124(10) & 0.13829(13) & 1.8144(62) & 0.24119(96)\\
 & 15 & 44  & 0.85813(13) & 0.13897(19) & 2.0195(68) & 0.2423(10)\\
 & 15 & 44  & 0.93926(16) & 0.13817(24) & 2.2109(74) & 0.2408(11)\\
\hline
\multirow{4}{*}{M1} & 13 & 32  & 0.67420(41) & 0.13560(46) & 1.6068(66) & 0.2407(13)\\
 & 13 & 32  & 0.76723(47) & 0.13774(64) & 1.8285(73) & 0.2446(15)\\
 & 13 & 32  & 0.85376(57) & 0.13800(86) & 2.0348(81) & 0.2450(19)\\
 & 13 & 32  & 0.93420(71) & 0.1365(12) & 2.2265(88) & 0.2423(23)\\
\hline
\multirow{4}{*}{M2} & 12 & 32  & 0.67472(41) & 0.13623(41) & 1.6081(66) & 0.2419(13)\\
 & 12 & 32  & 0.76794(50) & 0.13866(56) & 1.8302(73) & 0.2463(14)\\
 & 12 & 32  & 0.85470(84) & 0.1393(11) & 2.0370(83) & 0.2473(23)\\
 & 12 & 32  & 0.93539(64) & 0.13812(74) & 2.2293(88) & 0.2453(17)\\
\hline
\hline
\multirow{5}{*}{F1} & 14 & 43  & 0.57222(21) & 0.11343(20) & 1.5867(67) & 0.2393(12)\\
 & 14 & 43  & 0.65271(24) & 0.11546(26) & 1.8100(75) & 0.2436(12)\\
 & 14 & 43  & 0.72795(28) & 0.11611(34) & 2.0188(82) & 0.2450(13)\\
 & 14 & 43  & 0.79862(33) & 0.11562(43) & 2.2148(89) & 0.2439(15)\\
 & 14 & 43  & 0.89021(44) & 0.11320(61) & 2.4689(99) & 0.2388(17)\\
\hline
\end{tabular}
\caption{Fit results for the masses and decay constants of the $D_s$ meson on all ensembles. The second set of columns is renormalised where the renormalisation constants are obtained from the light-light conserved current as described in the text. The results stated here are those for the strange quark mass closest to the physical one.}
\label{tab:Ds_dynamical}
\end{table}

\begin{table}
\center
\small
\begin{tabular}{|l|rr||ll|}
\hline
ens & \multicolumn{2}{c||}{$[t_\mathrm{min},t_\mathrm{max})$} & $am_{\eta_c}$ & $m_{\eta_c}\,[\mathrm{GeV}]$ \\
\hline\hline
\multirow{3}{*}{C0} & 24 & 48  & 1.249409(56) & 2.1609(47)\\
 & 24 & 48  & 1.375320(51) & 2.3786(51)\\
 & 30 & 48  & 1.493579(48) & 2.5831(56)\\
\hline
\multirow{3}{*}{C1} & 18 & 32  & 1.24641(20) & 2.2246(62)\\
 & 21 & 32  & 1.37227(19) & 2.4492(68)\\
 & 21 & 32  & 1.49059(17) & 2.6604(74)\\
\hline
\multirow{3}{*}{C2} & 20 & 32  & 1.24701(20) & 2.2257(62)\\
 & 24 & 32  & 1.37276(18) & 2.4501(68)\\
 & 24 & 32  & 1.49102(16) & 2.6612(74)\\
\hline
\hline
\multirow{4}{*}{M0} & 28 & 59  & 0.972488(49) & 2.2937(70)\\
 & 36 & 59  & 1.135329(46) & 2.6778(81)\\
 & 38 & 59  & 1.287084(43) & 3.0357(92)\\
 & 38 & 59  & 1.428269(40) & 3.369(10)\\
\hline
\multirow{4}{*}{M1} & 22 & 32  & 0.96975(18) & 2.3112(83)\\
 & 23 & 32  & 1.13226(15) & 2.6985(97)\\
 & 23 & 32  & 1.28347(13) & 3.059(11)\\
 & 25 & 32  & 1.42374(12) & 3.393(12)\\
\hline
\multirow{4}{*}{M2} & 22 & 32  & 0.97000(19) & 2.3118(83)\\
 & 24 & 32  & 1.13246(18) & 2.6990(97)\\
 & 24 & 32  & 1.28365(16) & 3.059(11)\\
 & 26 & 32  & 1.42384(15) & 3.393(12)\\
\hline
\hline
\multirow{5}{*}{F1} & 31 & 48  & 0.82322(10) & 2.2836(83)\\
 & 31 & 48  & 0.965045(93) & 2.6770(98)\\
 & 31 & 48  & 1.098129(86) & 3.046(11)\\
 & 31 & 48  & 1.223360(80) & 3.394(12)\\
 & 35 & 48  & 1.385711(74) & 3.844(14)\\
\hline
\end{tabular}
\caption{Fit results for the masses of the connected part of the $\eta_c$ meson on all ensembles.}
\label{tab:etac_dynamical}
\end{table}

\section{Global fit results for $f_{D_s}/f_D$, $\Phi_D$ and $\Phi_{D_s}$}\label{app:GlobalFitResults}
\begin{table}[h]
  \center
      {\footnotesize
        \begin{tabular}{|cc||c||cc||cc|c|cc|}
          \hline
          $H$ & $m_\pi^\mathrm{cut}$ & $f_{D_s}/f_D$ & $C_\mathrm{CL}^0$ & $C_\mathrm{CL}^1$ & $C_\chi^0$ & $C_\chi^1$ & $C_h^0$ & $\chi^2/\mathrm{dof}$ & $p$\\
          \hline
          $\eta_c$ & 450 & 1.1531(60) & 0.037(22) & -  & -0.529(24) & -  & -0.022(16) & 0.739 & 0.803\\
          $\eta_c$ & 450 & 1.1526(72) & 0.036(23) & -  & -0.523(59) & -0.03(26) & -0.020(28) & 0.773 & 0.756\\
          $\eta_c$ & 450 & 1.157(11) & 0.019(44) & 0.10(21) & -0.527(25) & -  & -0.048(56) & 0.762 & 0.769\\
          $\eta_c$ & 450 & 1.157(11) & 0.016(46) & 0.11(21) & -0.515(61) & -0.06(26) & -0.045(57) & 0.798 & 0.719\\
          \hline
          ${D_s}$ & 450 & 1.1532(60) & 0.036(22) & -  & -0.529(24) & -  & -0.019(14) & 0.737 & 0.806\\
          ${D_s}$ & 450 & 1.1528(72) & 0.036(23) & -  & -0.523(58) & -0.02(22) & -0.017(24) & 0.771 & 0.759\\
          ${D_s}$ & 450 & 1.158(11) & 0.018(44) & 0.08(17) & -0.527(25) & -  & -0.041(47) & 0.760 & 0.772\\
          ${D_s}$ & 450 & 1.157(11) & 0.016(46) & 0.09(18) & -0.516(61) & -0.05(22) & -0.038(48) & 0.796 & 0.722\\
          \hline
          $D$ & 450 & 1.1531(60) & 0.036(22) & -  & -0.531(25) & -  & -0.016(12) & 0.740 & 0.801\\
          $D$ & 450 & 1.1521(70) & 0.035(23) & -  & -0.516(56) & -0.06(19) & -0.011(20) & 0.772 & 0.758\\
          $D$ & 450 & 1.156(11) & 0.023(43) & 0.06(15) & -0.530(25) & -  & -0.031(40) & 0.769 & 0.762\\
          $D$ & 450 & 1.155(11) & 0.019(44) & 0.06(15) & -0.510(58) & -0.07(19) & -0.027(41) & 0.801 & 0.716\\
          \hline
          \hline
          $\eta_c$ & 400 & 1.1667(77) & 0.005(25) & -  & -0.631(44) & -  & -0.031(19) & 0.319 & 0.998\\
          $\eta_c$ & 400 & 1.1644(90) & 0.003(25) & -  & -0.591(94) & -0.21(43) & -0.019(31) & 0.324 & 0.997\\
          $\eta_c$ & 400 & 1.171(12) & -0.015(51) & 0.11(23) & -0.628(45) & -  & -0.056(57) & 0.324 & 0.997\\
          $\eta_c$ & 400 & 1.168(14) & -0.012(51) & 0.08(24) & -0.597(95) & -0.16(44) & -0.041(69) & 0.335 & 0.995\\
          \hline
          ${D_s}$ & 400 & 1.1669(77) & 0.004(25) & -  & -0.631(44) & -  & -0.026(16) & 0.314 & 0.998\\
          ${D_s}$ & 400 & 1.1647(91) & 0.003(25) & -  & -0.592(94) & -0.17(36) & -0.016(27) & 0.319 & 0.997\\
          ${D_s}$ & 400 & 1.171(12) & -0.015(50) & 0.09(19) & -0.628(45) & -  & -0.047(48) & 0.320 & 0.997\\
          ${D_s}$ & 400 & 1.168(14) & -0.012(51) & 0.07(20) & -0.598(94) & -0.13(37) & -0.034(58) & 0.332 & 0.995\\
          \hline
          $D$ & 400 & 1.1668(78) & 0.004(25) & -  & -0.634(45) & -  & -0.022(14) & 0.319 & 0.998\\
          $D$ & 400 & 1.1644(89) & 0.002(25) & -  & -0.592(91) & -0.16(31) & -0.013(22) & 0.321 & 0.997\\
          $D$ & 400 & 1.171(12) & -0.014(48) & 0.07(16) & -0.631(46) & -  & -0.040(42) & 0.325 & 0.997\\
          $D$ & 400 & 1.168(14) & -0.010(49) & 0.05(17) & -0.598(92) & -0.13(32) & -0.027(50) & 0.335 & 0.995\\
          \hline
          \hline
          $\eta_c$ & 350 & 1.1655(79) & 0.006(26) & -  & -0.636(55) & -  & -0.025(22) & 0.352 & 0.989\\
          $\eta_c$ & 350 & 1.1653(92) & 0.006(26) & -  & -0.63(12) & -0.03(54) & -0.024(32) & 0.377 & 0.982\\
          $\eta_c$ & 350 & 1.168(13) & -0.005(53) & 0.06(25) & -0.635(55) & -  & -0.039(65) & 0.373 & 0.982\\
          $\eta_c$ & 350 & 1.168(14) & -0.005(53) & 0.06(25) & -0.63(12) & -0.01(54) & -0.039(70) & 0.402 & 0.970\\
          \hline
          ${D_s}$ & 350 & 1.1657(79) & 0.005(26) & -  & -0.637(55) & -  & -0.022(18) & 0.349 & 0.990\\
          ${D_s}$ & 350 & 1.1655(92) & 0.005(26) & -  & -0.63(12) & -0.02(45) & -0.021(28) & 0.374 & 0.982\\
          ${D_s}$ & 350 & 1.168(13) & -0.004(53) & 0.04(21) & -0.635(55) & -  & -0.032(55) & 0.371 & 0.983\\
          ${D_s}$ & 350 & 1.168(14) & -0.004(53) & 0.04(21) & -0.63(12) & -0.01(45) & -0.032(59) & 0.400 & 0.971\\
          \hline
          $D$ & 350 & 1.1656(79) & 0.005(26) & -  & -0.639(56) & -  & -0.018(16) & 0.352 & 0.989\\
          $D$ & 350 & 1.1652(90) & 0.005(26) & -  & -0.63(12) & -0.04(39) & -0.017(23) & 0.377 & 0.982\\
          $D$ & 350 & 1.168(13) & -0.003(52) & 0.03(18) & -0.638(56) & -  & -0.027(47) & 0.375 & 0.982\\
          $D$ & 350 & 1.167(14) & -0.003(52) & 0.03(18) & -0.63(12) & -0.03(39) & -0.025(50) & 0.403 & 0.970\\
          \hline
          \hline
      \end{tabular}}
      \caption{Fit results of the different versions of the global fit ansatz \eqref{eq:globalchiCL} for the observable $f_{D_s}/f_D$}
      \label{tab:global_fits_fDsfD}
\end{table}

\begin{table}
\center
{\footnotesize
\begin{tabular}{|cc||c||cc||cc|c|c|}
\hline
 $H$ & $m_\pi^\mathrm{cut}$ & $\Phi_D[\mathrm{GeV}^{3/2}]$ & $C_\mathrm{CL}^0$ & $C_\mathrm{CL}^1$ & $C_\chi^0$ & $C_\chi^1$ & $C_h^0$ & $\chi^2/\mathrm{dof}$ \\
\hline
$\eta_c$ & 450 & 0.2885(32) & -0.010(10) & -  & 0.203(13) & -  & -0.3797(86) & 0.555\\
$\eta_c$ & 450 & 0.2867(36) & -0.012(10) & -  & 0.234(26) & -0.36(21) & -0.354(18) & 0.459\\
$\eta_c$ & 450 & 0.2891(43) & -0.014(17) & 0.05(11) & 0.204(13) & -  & -0.390(30) & 0.577\\
$\eta_c$ & 450 & 0.2880(44) & -0.021(17) & 0.11(12) & 0.237(26) & -0.39(21) & -0.374(31) & 0.460\\
\hline
${D_s}$ & 450 & 0.2882(31) & -0.018(10) & -  & 0.200(12) & -  & -0.3061(70) & 0.540\\
${D_s}$ & 450 & 0.2866(34) & -0.020(10) & -  & 0.228(23) & -0.29(16) & -0.285(15) & 0.447\\
${D_s}$ & 450 & 0.2883(41) & -0.019(16) & 0.009(91) & 0.201(12) & -  & -0.308(24) & 0.565\\
${D_s}$ & 450 & 0.2874(42) & -0.025(16) & 0.056(94) & 0.230(24) & -0.31(17) & -0.296(25) & 0.460\\
\hline
$D$ & 450 & 0.2884(30) & -0.0202(95) & -  & 0.176(11) & -  & -0.2717(59) & 0.578\\
$D$ & 450 & 0.2868(32) & -0.0228(96) & -  & 0.205(21) & -0.30(13) & -0.251(12) & 0.446\\
$D$ & 450 & 0.2891(38) & -0.024(14) & 0.043(77) & 0.177(12) & -  & -0.281(20) & 0.598\\
$D$ & 450 & 0.2880(39) & -0.030(15) & 0.081(80) & 0.209(22) & -0.32(14) & -0.267(21) & 0.443\\
\hline
\hline
$\eta_c$ & 400 & 0.2853(38) & -0.003(11) & -  & 0.230(22) & -  & -0.3747(97) & 0.300\\
$\eta_c$ & 400 & 0.2841(43) & -0.005(11) & -  & 0.255(40) & -0.32(33) & -0.356(22) & 0.255\\
$\eta_c$ & 400 & 0.2876(48) & -0.017(18) & 0.17(13) & 0.233(23) & -  & -0.409(32) & 0.269\\
$\eta_c$ & 400 & 0.2861(55) & -0.015(19) & 0.14(14) & 0.254(40) & -0.28(34) & -0.386(44) & 0.237\\
\hline
${D_s}$ & 400 & 0.2851(37) & -0.011(11) & -  & 0.227(22) & -  & -0.3016(79) & 0.279\\
${D_s}$ & 400 & 0.2840(42) & -0.012(11) & -  & 0.248(37) & -0.25(26) & -0.288(17) & 0.241\\
${D_s}$ & 400 & 0.2867(46) & -0.021(17) & 0.11(10) & 0.230(22) & -  & -0.323(25) & 0.265\\
${D_s}$ & 400 & 0.2853(52) & -0.020(18) & 0.08(11) & 0.248(37) & -0.22(27) & -0.306(35) & 0.237\\
\hline
$D$ & 400 & 0.2854(36) & -0.014(11) & -  & 0.203(20) & -  & -0.2670(66) & 0.316\\
$D$ & 400 & 0.2840(39) & -0.015(11) & -  & 0.230(34) & -0.30(22) & -0.250(14) & 0.232\\
$D$ & 400 & 0.2875(43) & -0.026(16) & 0.135(87) & 0.207(21) & -  & -0.294(22) & 0.271\\
$D$ & 400 & 0.2857(48) & -0.025(16) & 0.103(92) & 0.229(34) & -0.26(22) & -0.273(29) & 0.209\\
\hline
\hline
$\eta_c$ & 350 & 0.2855(39) & -0.005(12) & -  & 0.235(25) & -  & -0.372(11) & 0.348\\
$\eta_c$ & 350 & 0.2835(43) & -0.006(12) & -  & 0.281(48) & -0.56(40) & -0.346(22) & 0.238\\
$\eta_c$ & 350 & 0.2878(50) & -0.018(19) & 0.17(14) & 0.239(25) & -  & -0.407(35) & 0.312\\
$\eta_c$ & 350 & 0.2856(55) & -0.017(19) & 0.14(14) & 0.281(48) & -0.53(40) & -0.378(43) & 0.209\\
\hline
${D_s}$ & 350 & 0.2852(38) & -0.013(11) & -  & 0.232(24) & -  & -0.2996(87) & 0.319\\
${D_s}$ & 350 & 0.2835(42) & -0.014(11) & -  & 0.272(45) & -0.43(31) & -0.280(17) & 0.225\\
${D_s}$ & 350 & 0.2868(47) & -0.022(18) & 0.11(11) & 0.235(25) & -  & -0.322(28) & 0.304\\
${D_s}$ & 350 & 0.2850(52) & -0.022(18) & 0.09(11) & 0.272(45) & -0.41(32) & -0.299(34) & 0.214\\
\hline
$D$ & 350 & 0.2854(36) & -0.015(11) & -  & 0.208(23) & -  & -0.2639(73) & 0.352\\
$D$ & 350 & 0.2836(39) & -0.017(11) & -  & 0.252(40) & -0.45(25) & -0.244(14) & 0.209\\
$D$ & 350 & 0.2876(44) & -0.028(17) & 0.134(93) & 0.213(23) & -  & -0.292(24) & 0.300\\
$D$ & 350 & 0.2855(48) & -0.028(17) & 0.111(93) & 0.253(40) & -0.42(26) & -0.268(28) & 0.169\\
\hline
\hline
    \end{tabular}}
\caption{Fit results of the different versions of the global fit ansatz \eqref{eq:globalchiCL} for the observable $\Phi_D$}
\label{tab:global_fits_fmD}
\end{table}

\begin{table}
\center
{\footnotesize
\begin{tabular}{|cc||c||cc||cc|c|c|}
\hline
 $H$ & $m_\pi^\mathrm{cut}$ & $\Phi_{D_s}[\mathrm{GeV}^{3/2}]$ & $C_\mathrm{CL}^0$ & $C_\mathrm{CL}^1$ & $C_\chi^0$ & $C_\chi^1$ & $C_h^0$ & $\chi^2/\mathrm{dof}$ \\
\hline
$\eta_c$ & 450 & 0.3449(21) & -0.0206(68) & -  & 0.0655(92) & -  & -0.4167(38) & 0.297\\
$\eta_c$ & 450 & 0.3448(21) & -0.0206(69) & -  & 0.066(16) & -0.01(14) & -0.4162(76) & 0.311\\
$\eta_c$ & 450 & 0.3463(26) & -0.0286(98) & 0.114(65) & 0.0665(92) & -  & -0.440(15) & 0.262\\
$\eta_c$ & 450 & 0.3462(25) & -0.0287(100) & 0.115(66) & 0.068(16) & -0.02(14) & -0.440(16) & 0.274\\
\hline
${D_s}$ & 450 & 0.3444(18) & -0.0294(59) & -  & 0.0622(80) & -  & -0.3345(30) & 0.261\\
${D_s}$ & 450 & 0.3444(19) & -0.0294(60) & -  & 0.064(14) & -0.02(10) & -0.3337(56) & 0.273\\
${D_s}$ & 450 & 0.3451(23) & -0.0334(87) & 0.051(50) & 0.0628(80) & -  & -0.345(12) & 0.258\\
${D_s}$ & 450 & 0.3451(23) & -0.0336(88) & 0.052(50) & 0.065(14) & -0.02(10) & -0.344(12) & 0.270\\
\hline
$D$ & 450 & 0.3448(20) & -0.0327(66) & -  & 0.0364(88) & -  & -0.2952(27) & 0.359\\
$D$ & 450 & 0.3446(20) & -0.0331(67) & -  & 0.044(13) & -0.086(91) & -0.2914(50) & 0.349\\
$D$ & 450 & 0.3459(24) & -0.0388(91) & 0.073(45) & 0.0378(88) & -  & -0.311(11) & 0.336\\
$D$ & 450 & 0.3456(24) & -0.0394(92) & 0.074(45) & 0.045(13) & -0.089(91) & -0.307(10) & 0.323\\
\hline
\hline
$\eta_c$ & 400 & 0.3442(22) & -0.0185(68) & -  & 0.073(14) & -  & -0.4175(40) & 0.278\\
$\eta_c$ & 400 & 0.3442(22) & -0.0185(69) & -  & 0.073(21) & 0.01(18) & -0.4179(77) & 0.293\\
$\eta_c$ & 400 & 0.3457(26) & -0.027(10) & 0.122(70) & 0.075(14) & -  & -0.442(16) & 0.235\\
$\eta_c$ & 400 & 0.3459(27) & -0.028(10) & 0.127(71) & 0.072(21) & 0.05(18) & -0.445(18) & 0.246\\
\hline
${D_s}$ & 400 & 0.3438(19) & -0.0275(59) & -  & 0.070(12) & -  & -0.3348(32) & 0.243\\
${D_s}$ & 400 & 0.3438(19) & -0.0274(60) & -  & 0.069(19) & 0.01(13) & -0.3353(57) & 0.256\\
${D_s}$ & 400 & 0.3446(24) & -0.0322(90) & 0.058(53) & 0.071(12) & -  & -0.347(12) & 0.236\\
${D_s}$ & 400 & 0.3447(24) & -0.0324(90) & 0.062(54) & 0.069(19) & 0.03(13) & -0.348(14) & 0.248\\
\hline
$D$ & 400 & 0.3442(21) & -0.0310(66) & -  & 0.044(13) & -  & -0.2951(28) & 0.358\\
$D$ & 400 & 0.3440(21) & -0.0314(66) & -  & 0.050(19) & -0.08(12) & -0.2920(51) & 0.359\\
$D$ & 400 & 0.3455(25) & -0.0386(93) & 0.087(48) & 0.046(13) & -  & -0.313(11) & 0.318\\
$D$ & 400 & 0.3453(25) & -0.0382(92) & 0.080(47) & 0.049(19) & -0.05(12) & -0.309(12) & 0.328\\
\hline
\hline
$\eta_c$ & 350 & 0.3444(22) & -0.0202(69) & -  & 0.079(15) & -  & -0.4152(42) & 0.268\\
$\eta_c$ & 350 & 0.3445(22) & -0.0201(69) & -  & 0.075(26) & 0.06(23) & -0.4167(72) & 0.284\\
$\eta_c$ & 350 & 0.3458(27) & -0.028(10) & 0.113(72) & 0.080(15) & -  & -0.438(17) & 0.224\\
$\eta_c$ & 350 & 0.3461(27) & -0.029(10) & 0.119(68) & 0.073(26) & 0.09(22) & -0.442(16) & 0.232\\
\hline
${D_s}$ & 350 & 0.3440(19) & -0.0290(60) & -  & 0.075(13) & -  & -0.3331(34) & 0.227\\
${D_s}$ & 350 & 0.3441(19) & -0.0290(60) & -  & 0.072(22) & 0.04(17) & -0.3344(54) & 0.240\\
${D_s}$ & 350 & 0.3447(24) & -0.0332(91) & 0.052(54) & 0.076(13) & -  & -0.344(13) & 0.223\\
${D_s}$ & 350 & 0.3449(24) & -0.0334(90) & 0.056(52) & 0.071(22) & 0.06(16) & -0.346(12) & 0.234\\
\hline
$D$ & 350 & 0.3445(21) & -0.0330(67) & -  & 0.049(14) & -  & -0.2924(30) & 0.326\\
$D$ & 350 & 0.3444(21) & -0.0331(67) & -  & 0.053(22) & -0.05(15) & -0.2911(48) & 0.346\\
$D$ & 350 & 0.3456(25) & -0.0398(94) & 0.079(49) & 0.051(14) & -  & -0.309(12) & 0.288\\
$D$ & 350 & 0.3455(25) & -0.0397(93) & 0.077(46) & 0.052(22) & -0.02(15) & -0.308(11) & 0.309\\
\hline
\hline
    \end{tabular}}
\caption{Fit results of the different versions of the global fit ansatz \eqref{eq:globalchiCL} for the observable $\Phi_{D_s}$}
\label{tab:global_fits_fmDs}
\end{table}

\clearpage
\section{RI/SMOM and the axial vertex function}\label{app:nprsmom}
The projected axial vertex functions are generated according to the SMOM renormalization condition~\cite{Sturm:2009kb}:
\eq{
  \label{eq:SMOM-axial}
  \lim_{\overline{m}\to 0}  \frac{1}{12 q^2} \text{Tr} \left. \left[ \left(q
    \cdot \Lambda_{\text{A},R} \right) \gamma_5 \slashed{q} \right]
  \right|_\text{sym} = 1 ,
}
where $\Lambda_{\text{A},R}$ is the amputated axial vertex function and the subscript $R$ denotes a renormalised quantity.
The momentum $q$ out of the vertex satisfies the symmetric non-exceptional condition:
\eq{
  p_1^2 =p_2^2 =q^2 .
} 
\begin{figure}
  \centering
  \includegraphics[scale=0.5]{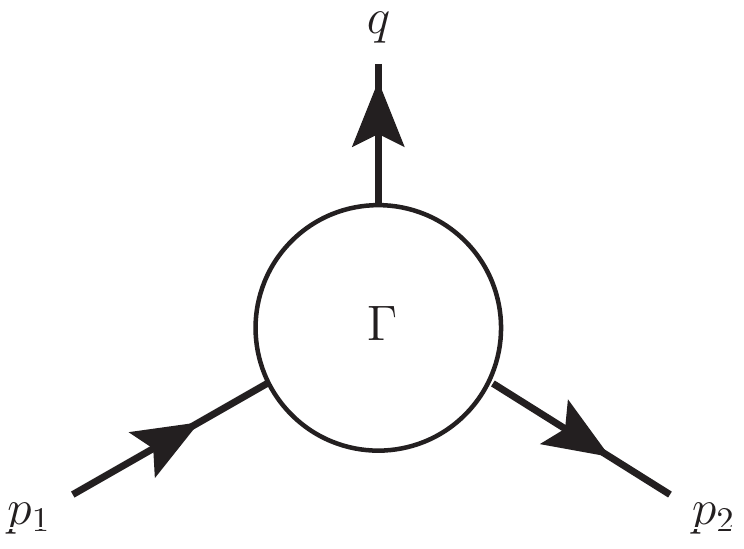}
  \caption{Kinematics used for the correlators of fermion bilinears. Here we have $\Gamma=\text{A}=\gamma^\mu\gamma^5$}
  \label{fig:kin}
\end{figure}
The momenta are determined by 
\eq{
  ap_\mu=n_\mu\frac{2\pi}{L_\mu/a} \ ,
}
for every lattice spacing $L$ such that the magnitude of $p$ is around $2\,\mathrm{GeV}$ for an integer $n$.
Note that in order to reach the intermediate momenta we use twisting~\cite{Arthur:2010ht}:
\eq{
  ap_\mu=\left(n_\mu+\frac{\theta_\mu}{2}\right)\frac{2\pi}{L_\mu/a} \ .
}

Eq. \eqref{eq:SMOM-axial} can be written in terms of the amputated bare vertex function, the field renormalisation $Z_q$ and the axial operator renormalisation $Z_A$ as follows,
\eq{
  \label{eq:SMOM-axial-bare}
  \lim_{\overline{m}\to 0}  \frac{1}{12 q^2} \frac{Z_\text{A}}{Z_q}\text{Tr} \left. \left[ \left(q
    \cdot \Lambda_{\text{A}} \right) \gamma_5 \slashed{q} \right]
  \right|_\text{sym} = 1 ,
}
where the bare quantity is what is computed numerically on the lattice.
Above we have denoted 
\eq{
  \mathcal{P}[\Lambda_A] \equiv \lim_{\overline{m}\to 0}  \frac{1}{12 q^2} \text{Tr} \left. \left[ \left(q
    \cdot \Lambda_{\text{A}} \right) \gamma_5 \slashed{q} \right]
  \right|_\text{sym} . 
}
In principle, if the action is substantially changed for the heavy as compared to light quarks, it is possible in the massless renormalisation scheme, to apply the axial RI-SMOM condition \eqref{eq:SMOM-axial} to the mixed action bilinear
\eq{
\mathcal{P}[\Lambda_\text{A}](1.6,1.8) = \frac{\sqrt{Z_q^1}\sqrt{Z_q^2}}{Z_A^{12}}.
}
Again, the indices $i=1,2$ refer to the action of the first and second quark field entering the bilinear operator. In the chiral limit, it is possible to systematically eliminate the factors
of the quark field renormalisation $Z_q$ from the SMOM condition~\cite{Boyle:2016wis}. However, since we can compute the corresponding unmixed vertex functions, and know how to determine $Z_A^{11}$ and $Z_A^{22}$
from the conserved domain wall current as in eq.~\eqref{eq:ZAeff}, we are able to determine the axial current renormalisation, $Z_A^{12}$,
from ratios of vertex functions as follows:
\eq{
\frac{\left(\mathcal{P}[\Lambda_\text{A}](1.6,1.6)\right)
 \left(\mathcal{P}[\Lambda_\text{A}](1.8,1.8)\right)}
{\left(\mathcal{P}[\Lambda_\text{A}](1.6,1.8)\right)^2}
 = \frac{\left(Z_A^{12}\right)^2}{Z_A^{11}Z_A^{22}}.
}
This result is an interesting, and fully non-perturbative, analogue of the Fermilab~\cite{Hashimoto:1999yp}
partially-perturbative approach to currents that are conserved in the continuum. The ratio of vertex functions in our case is very near unity since the difference in the actions of the quark legs is very small, as is seen in tables \ref{tab:lambdaAratiosC2}, \ref{tab:lambdaAratiosM1} and \ref{tab:lambdaAratiosF1}.

Furthermore, if the RI-SMOM conditions are modified we can form a consistent set of conditions in the massive case~\cite{Boyle:2016wis}. Since the axial current is partially conserved in the continuum, the difference between the schemes is necessarily only a lattice artefact, implying that either approach may be taken when we take the continuum limit, and yields a universal result. The scaling violations will of course differ between these approaches and for the present paper we have taken the simpler approach of using the near massless vertex function data to define the scaling trajectory for axial current matrix elements.

\clearpage

{\small
\bibliographystyle{JHEP}    

\bibliography{charm}}

\end{document}